\documentclass[10pt,aps,pra,twocolumn,floatfix,nofootinbib,superscriptaddress,longbibliography]{revtex4-2}
\usepackage[utf8]{inputenc}

\usepackage[dvipsnames]{xcolor}
\definecolor{alecolor}{RGB}{198,113,190}
\usepackage{bm,graphicx,mathrsfs,amsmath,amssymb,mathtools,makecell,bbm,amsthm,dsfont,color,times,txfonts,nicefrac,framed,enumitem,tikz,physics,wrapfig,amsfonts,lipsum,nicematrix,xargs, multirow}
\definecolor{equationcolor}{RGB}{222,94,100}
\definecolor{changescolor}{rgb}{0, 0, 0.7}
\newcommand{\iden}{\mathbbm{1}}
\usepackage[colorlinks=true,linkcolor=alecolor,citecolor=equationcolor]{hyperref}

\hypersetup{urlcolor=equationcolor}

\definecolor{equationcolor}{RGB}{154,135,198}
\usetikzlibrary{arrows.meta}
\usetikzlibrary{positioning}
\usetikzlibrary{calc} 
\usepackage{comment}

\begin{document}

\title{Semi–device-independent randomness certification on discretized continuous-variable platforms}

\author{Moisés Alves}
\affiliation{Physics Department, Federal University of Rio Grande do Norte, Natal, 59072-970, Rio Grande do Norte, Brazil}
\affiliation{International Institute of Physics, Federal University of Rio Grande do Norte, 59078-970, Natal, Brazil}
\affiliation{QuIIN - Quantum Industrial Innovation, EMBRAPII CIMATEC Competence Center in Quantum Technologies, SENAI CIMATEC, Av. Orlando Gomes 1845, 41650-010, Salvador, BA, Brazil. }
\author{Vitor L. Sena}
\affiliation{QuIIN - Quantum Industrial Innovation, EMBRAPII CIMATEC Competence Center in Quantum Technologies, SENAI CIMATEC, Av. Orlando Gomes 1845, 41650-010, Salvador, BA, Brazil. }
\author{Santiago Zamora}
\affiliation{Physics Department, Federal University of Rio Grande do Norte, Natal, 59072-970, Rio Grande do Norte, Brazil}
\affiliation{International Institute of Physics, Federal University of Rio Grande do Norte, 59078-970, Natal, Brazil}
\author{Tailan S. Sarubi}
\affiliation{Physics Department, Federal University of Rio Grande do Norte, Natal, 59072-970, Rio Grande do Norte, Brazil}
\affiliation{International Institute of Physics, Federal University of Rio Grande do Norte, 59078-970, Natal, Brazil}
\author{A. de Oliveira Junior}
\affiliation{Center for Macroscopic Quantum States bigQ, Department of Physics,
Technical University of Denmark, Fysikvej 307, 2800 Kgs. Lyngby, Denmark}
\author{Alexandre B. Tacla}
\affiliation{QuIIN - Quantum Industrial Innovation, EMBRAPII CIMATEC Competence Center in Quantum Technologies, SENAI CIMATEC, Av. Orlando Gomes 1845, 41650-010, Salvador, BA, Brazil. }
\author{Rafael Chaves}
\affiliation{International Institute of Physics, Federal University of Rio Grande do Norte, 59078-970, Natal, Brazil}
\affiliation{School of Science and Technology, Federal University of Rio Grande do Norte, Natal, Brazil}

\begin{abstract}
Randomness is fundamental for secure communication and information processing. While continuous-variable optical systems offer an attractive platform for this task, certifying genuine quantum randomness in such setups remains challenging. We present a semi–device–independent scheme for randomness certification tailored to continuous-variable implementations, where the dimension assumption is operationally implemented by restricting state preparations to the two-level Fock subspace. Using standard homodyne and displacement-based measurements, we show that simple optical setups can achieve dimension-witness violations that certify positive min-entropy, even in the presence of realistic losses and misaligned reference frames. These results demonstrate that practical and scalable quantum randomness generation is achievable with minimal experimental complexity on continuous-variable platforms.
\end{abstract}

\maketitle

\section{Introduction}
Randomness has become a cornerstone of modern science and technology. A plethora of celebrated examples ranges from computational methods~\cite{metropolis1949monte,motwani1996randomized,rabin1980probabilistic} to cryptographic security~\cite{blum1986simple,shannon1949communication}, as well as many other facets of quantum theory~\cite{Pironio2010,Herrero2017}. Classical devices, however, can only implement pseudorandomness -- deterministic processes that mimic chance -- whereas quantum measurements exhibit intrinsic unpredictability that cannot be explained by hidden variables~\cite{Bell1964,Kochen1975}. This unique feature forms the basis of quantum random-number generators (QRNGs)~\cite{Herrero2017}. Yet, using a quantum device alone does not guarantee that the generated bits are truly unpredictable to an adversary, as imperfections or implementation flaws can still produce apparent randomness~\cite{Jennewein2000,Pironio2010,Acin2016,Herrero2017}.

\begin{figure}[t]
    \centering
    \includegraphics{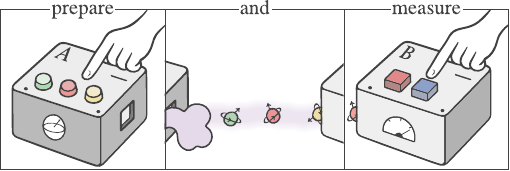}
    \caption{\textbf{Prepare-and-measure scenario.} Alice’s device prepares a quantum state $\rho_x$ upon pressing a coloured button $x\in\{\text{green},\text{red},\text{yellow}\}$ and sends the system to Bob. In his lab, Bob chooses a measurement labeled by $y\in\{\text{blue},\text{red}\}$ and records an outcome $b$. By repeating the procedure many times, the parties estimate the conditional statistics $p(b| x,y)$, which are later analyzed to assess their (non)classical nature and the presence of certifiable randomness.
   }
    \label{F:PAM-scenario}
\end{figure}

Certification frameworks span a spectrum of trust assumptions. At one extreme, device‑independent (DI) protocols certify randomness from a Bell‑inequality violation without modeling the devices’ internals, but they require highly demanding experimental conditions, such as loophole‑free detection with spacelike separation~\cite{hensen2015loophole,Giustina2015,Zhao2024}. Relaxing assumptions leads to semi‑device‑independent (SDI) approaches~\cite{Gallego2010,Brunner2011,Li2011,Brunner2013,Pauwels2022,Pauwels2025informationcapacity}, where one constrains a resource that limits any classical simulation. A powerful SDI setting is the prepare-and-measure (PAM) scenario with a dimension bound~(see Fig.~\ref{F:PAM-scenario} for a pictorial representation): if the communicated system is of bounded dimension, then certain dimension witnesses -- linear functionals of the observed correlations -- separate classical from quantum behaviors~\cite{Gallego2010,Bowles2014,VanHimbeeck2017}. A violation of such a witness simultaneously rules out any classical model respecting the constraint and certifies a lower bound on the system’s quantum dimension. This provides the basis for SDI randomness certification, showing that the observed outcomes were not fixed in advance by any classical explanation consistent with the bound.

The strength of this certification can be quantified: for a given family of witnesses, larger violations translate into tighter limits on an adversary’s ability to guess the outcomes and, consequently, into more guaranteed random bits per trial~\cite{Li2011,Brunner2011,Acin2016}. Although witness‑to‑randomness methods are well developed in discrete‑variable settings, their continuous‑variable (CV) counterparts remain less explored despite the appeal of standard optical hardware~\cite{Weedbrook2012}. In particular, displacement‑based on/off photodetection and homodyne measurement are experimentally accessible and low‑loss, making them natural candidates for scalable SDI implementations~\cite{Wittmann2010}. Turning these advantages into certified randomness in a dimension‑bounded prepare‑and‑measure setting is, however, non‑trivial~\cite{Nha2004,Julien2009}. Homodyne produces continuous outcomes that must be binned to yield the discrete data used by dimension witnesses (whereas displacement is inherently binary); the communicated system’s effective dimension must be enforced in practice to rule out classical explanations; and the violation‑to‑randomness link must be established for CV data and shown robust to realistic imperfections such as finite detection efficiency \cite{dall2012robustness} and misaligned reference frames \cite{shadbolt2012guaranteed}.

We address these open problems by presenting a SDI randomness-certification scheme tailored to continuous-variable optics within a dimension-bounded prepare-and-measure scenario. Our scheme uses only experimentally accessible homodyne and displacement-based on/off detection and imposes the dimension assumption operationally by restricting preparations to the lowest two Fock levels, guaranteeing explicit control over the communicated system. We introduce a simple discretization of homodyne outcomes and establish an analytical witness-to-randomness mapping that converts observed violations into certified extractable randomness per trial. Benchmarking across homodyne-only, displacement-only, and hybrid PAM configurations gives optimal witness values and identifies the measurement choices that maximize certification. Prior SDI randomness certification in the dimension-bounded PAM setting has focused essentially on random-access-code (RAC) witnesses, and on extensions to larger input sets~\cite{Li2011,li2012semi,Li2015,rusca2019self}. To the best of our knowledge, linear witnesses for the simplest PAM scenarios, with three preparations and two or three measurements, have not previously been used to derive min-entropy bounds in the standard dimension-bounded PAM model. For the three-measurement configuration, we also determine the quantum bounds of two previously uncharacterized witnesses, filling a gap in the literature. Finally, we assess robustness under realistic imperfections by modeling finite detection efficiency and dispensing with shared reference frames, showing that witness violations ---and the associated certified randomness--- persist under losses and unknown relative phases.

Importantly, complementary SDI route replaces a dimension bound with a physically motivated constraint on the expectation value of a chosen observable, such as the energy/mean photon number. In this energy‑constrained PAM, one characterizes the attainable correlations under limits on photon‑number statistics and can certify randomness without modeling the devices’ internals. This idea was formalized in optics by bounding the average photon number~\cite{VanHimbeeck2017,VanHimbeeck2019Energy}, and it underlies several homodyne/heterodyne–based QRNG protocols and demonstrations, including self‑testing schemes under energy assumptions and security against quantum side information~\cite{Rusca2019SelfTesting,Rusca2020Homodyne,Avesani2021Heterodyne,Wang2023Homodyne}. Most recently, photon‑number–constrained PAM scenarios were analyzed with semidefinite relaxations, yielding practical performance bounds and randomness certification (via conditional Shannon‑entropy bounds) and improving extraction for coherent‑state–plus‑homodyne implementations~\cite{Roch2025}. These developments emphasize that, for CV optics, energy‑style assumptions offer an appealing and experimentally natural alternative to dimension bounds. Our work focuses on the dimension-bounded model, while the same measurement settings (homodyne and on/off displacement) are compatible with both frameworks.

The paper is organized as follows. Section~\ref{Sec:prepare-and-measure-scenario} introduces the semi–device–independent prepare-and-measure framework, the classical causal benchmark, the dimension-witness families used, the randomness metrics, and the continuous-variable measurement schemes (homodyne with binning and displacement-based on/off detection). Section~\ref{Sec:results} presents our main results. First, we compute optimal witness values for homodyne-only, displacement-only, and hybrid configurations in the scenarios with up to four preparations and up to three measurement setups and compare performance across these configurations. Second, we assess robustness to finite detection efficiency via a loss model (deriving critical thresholds) and demonstrate certification without shared reference frames by observing violations under unknown relative phases. Section~\ref{Sec:final-remarks} concludes. Technical details are deferred to Appendices~\ref{Sec:app-optimal-strategies} and~\ref{Sec:app-analytical-upper-bound}.

\section{Prepare-and-measure scenario \& methods}\label{Sec:prepare-and-measure-scenario}

\subsection{General framework}

We consider a prepare-and-measure scenario with two parties, Alice and Bob, defined by the following two steps (see Fig.~\ref{F:PAM-scenario} for a pictorial illustration):
\begin{enumerate}
    \item \emph{Preparation}: Alice receives a classical input $x \in \{1,\dots,n_x\}$ and prepares a quantum state $\rho_x$ on a $d$-dimensional Hilbert space $\mathcal{H}_d$. The system is then sent to Bob. 
    
    \item \emph{Measurement}: Upon receiving the system, Bob is given an independently chosen input $y \in \{1,\dots,n_y\}$ that determines the measurement $M_y=\{M_{b|y}\}_{b=0}^{n_b-1}$. Performing this measurement results in a classical outcome $b \in \{0,\dots,n_b-1\}$.
\end{enumerate}
The aim of the PAM scenario is to determine when the observed input-output behavior cannot be explained by a classical model. In particular, we ask under which operational assumptions -- such as a bound on the dimension of the communicated system and the absence of entanglement assistance -- the outcomes on Bob’s side contain genuine randomness that is unpredictable to any adversary consistent with those assumptions.

In what follows, we adopt a semi–device–independent viewpoint: the preparation and measurement devices are treated as black boxes whose internal workings are uncharacterized. The only assumptions we make are operational. First, the system communicated from $A$ to $B$ has a bounded dimension $d$. Second, the choice of measurement setting $y$ is independent of $x$ and of any internal variables of the devices (measurement independence). Third, within each run, the causal flow is one-way -- from $A$ to $B$ -- with no signaling from $B$ back to $A$. Finally, while the devices may share pre-established classical randomness, we exclude entanglement assistance \cite{Tavakoli2021,Pauwels2022,RochTavakoli2024,Pawlowski2010,moreno2021semi}. 

We consider a scenario specified by the tuple $(n_x,n_y,n_b,d)$, where $n_x$ and $n_y$ are the numbers of possible inputs for Alice and Bob, respectively, $n_b$ is the number of outcomes for each measurement, and $d$ is the Hilbert-space dimension of the communicated system. The behavior of the setup is characterized by the conditional distribution $p(b| x,y)$, i.e., the probability that Bob obtains outcome $b$ when Alice prepares $\rho_x$ and Bob implements $M_y=\{M_{b\mid y}\}_{b=0}^{n_b-1}$. According to the Born rule, these probabilities are given by
\begin{equation}\label{Eq:PAM-born-rule}
    p(b|x,y) = \tr(\rho_x M_{b|y}).
\end{equation} 
The central question we address is whether the observed statistics $p(b|x,y)$ can certify that the outcome $b$ is genuinely random. The unpredictability of measurement outcomes is closely linked to whether the observed probability distribution admits a classical hidden-variable explanation.

\begin{figure}[t]
    \centering
    \includegraphics{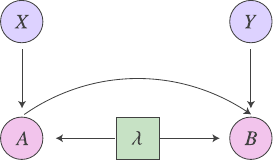}
    \caption{\textbf{Causal structure of the PAM scenario.} Inputs $X$ and $Y$ select the preparation device $A$ and the measurement device $B$. The devices may share pre-established classical randomness modelled by a latent variable $\lambda$ (arrows $\lambda \to A$ and $\lambda \to B$). Within each run the only communication is from $A$ to $B$ (edge $A \to B$, bounded to dimension $d$).
   }
    \label{F:PAM-dag}
\end{figure}

Following Bell’s paradigmatic example, classicality can be defined by imposing an underlying causal structure on the experiment. In this picture, quantum states and measurements are replaced by random variables with well-defined input–output relations. The preparation and measurement devices are then allowed to share pre-established correlations, represented by a hidden variable $\lambda$, which links their respective strategies. This classical scenario is captured by the causal structure shown in Fig.~\ref{F:PAM-dag}, implying that any probability distribution compatible with it must take the form~\cite{Gallego2010}
\begin{equation}\label{Eq:class-model}
    p(b\vert x,y)= \sum_{\lambda,a}p(b\vert a,y,\lambda) p(a\vert x,\lambda)p(\lambda).\
\end{equation}

Without additional constraints, all quantum correlations described by Eq.~\eqref{Eq:PAM-born-rule} can be reproduced by a classical model of the form~\eqref{Eq:class-model}. In particular, if the communicated system is unbounded (e.g., $d\ge n_x$ so that preparations are perfectly distinguishable), Alice can encode $x$ into either a set of $n_x$ orthogonal quantum states (when $d\ge n_x$) or a classical message with $\log_2 n_x$ bits, which Bob can perfectly read out. To avoid this trivialization we impose a dimension bound $d<n_x$, limiting the distinguishability of preparations per use. Related semi–device–independent assumptions used in practice constrain the $A\to B$ information in other ways: state-overlap constraints~\cite{brask2017megahertz,Wang2019}, minimum indistinguishability via fidelity bounds~\cite{Shi2019}, entropy constraints on preparations~\cite{chaves2015device}, energy bounds~\cite{VanHimbeeck2017,rusca2019self}, and direct limits expressed via (conditional) min-entropy~\cite{Tavakoli2019}. These are not identical to a hard dimension bound but have the same purpose of ruling out classical explanations.

\subsection{Dimension witnesses \& randomness metrics}
A non-classicality witness in the PAM scenario is any linear inequality of the form
\begin{equation}
    W = \sum_{b,x,y} W_{b|x,y}\;p(b|x,y) \leq W^C,
    \label{eq: GeneralW}
\end{equation}
where $W_{b|x,y} \in \mathbb{R}$ and $W^C$ is the classical bound. Any probability distribution $p(b|x,y)$ obtained from an experiment respecting the causal structure in Fig.~\ref{F:PAM-dag} satisfies Ineq.~\eqref{eq: GeneralW}. In quantum theory, one can achieve a larger value $W^Q > W^C$. Thus, a violation of Ineq.~\eqref{eq: GeneralW} certifies that the behavior $p(b|x,y)$ is non-classical, i.e., incompatible with the causal model~\eqref{Eq:class-model}.

We consider witnesses for the scenarios $(3,2,2,2)$, $(4,2,2,2)$, and $(3,3,2,2)$. All these scenarios have been studied in the discrete-variable PAM literature for nonclassicality and dimension testing~\cite{Gallego2010,Tavakoli2018,poderini2020criteria}. Randomness-certification analyses in the dimension-bounded PAM setting, however, have focused predominantly on the $(4,2,2,2)$ case tied to $2\!\to\!1$ QRACs (see, e.g.,~\cite{Li2011,rusca2019self,Mironowicz2014}). To the best of our knowledge, analogous min-entropy bounds for the linear witnesses in $(3,2,2,2)$ and $(3,3,2,2)$ facet families have not been reported. We will translate these witnesses to a discretized continuous-variable platform, realizing them with linear-optics measurements---homodyne, displacement, and hybrid schemes combining both---and evaluating performance under realistic imperfections (loss, finite efficiency, and phase misalignment). We begin by recalling the main ingredients of each scenario.

The $(3,2,2,2)$ case is the simplest non-trivial PAM scenario with a dimensional constraint. It is fully characterized by the inequality
\begin{equation}
    S_3 = E_{11} + E_{12} + E_{21} - E_{22} - E_{31} \le S_3^C = 3,
    \label{eq:dw_ineqS3}
\end{equation}
where $E_{xy} = p(0|x,y) - p(1|x,y)$ is the expectation value. This inequality defines a facet of the classical correlations polytope; all other facet-defining inequalities are obtained by relabeling the inputs $x,y$ or by sign permutations. Under the same dimensional constraint, quantum correlations can violate the $S_3$ inequality up to a maximum value $S_3^Q = 1 + 2\sqrt{2} \approx 3.82$~\cite{Gallego2010}. 

One can generalize the $S_3$ inequality to a one-parameter family that is sensitive to asymmetry in the $(3,2,2,2)$ configuration~\cite{certif_asymmetry} . A convenient ``tilted'' form reweights one of the correlators:
\begin{equation}
     S_3(w) = w(E_{11} + E_{21} - E_{31}) + (1-w)(E_{12} - E_{22}) \le S_3^C(w),
     \label{eq: s3_tilted}                          
\end{equation}
with classical bound $S_3^C(w) \!=\! \max(2-w, 3w)$ for $w \!\in\! [0,1]$. The corresponding quantum value is $S_3^Q(w) = 2\sqrt{w^2 + (1-w)^2} + w$ for $w \in [0,1]$. Note that $S_3\left(\tfrac12\right)=\tfrac12S_3$, i.e., the original facet is recovered up to an overall factor $2$.

The classical polytope of the $(4,2,2,2)$ scenario is described by two classes of facets. One is equivalent (up to relabellings/sign flips) to \eqref{eq:dw_ineqS3}. The other is,
\begin{align}
    S_4 = E_{11} &+ E_{12} + E_{21} - E_{22}\nonumber\\
    &-E_{31} + E_{32} - E_{41} - E_{42}\leq S_4^C = 4.
    \label{eq: s4}
\end{align}
The quantum maximum is $S_4^Q = 4\sqrt{2}$. This follows by observing that $S_4$ is the sum of two CHSH functionals (one on inputs $x\in\{1,2\})$, the other on $x\in\{3,4\}$), each achieving the Tsirelson bound $2\sqrt{2}$ with suitable settings --- hence the total $4\sqrt{2}$. Observe that an analogous reasoning recovers the quantum maximum of $2\sqrt{2} +1$ for $S_3$.

The last scenario we consider increases the number of measurements while keeping preparations $n_x = 3$. This $(3,3,2,2)$ case admits three facet classes \cite{poderini2020criteria}. The first one again corresponds to $S_3$, and the remaining two are given by
\begin{align}
    S_{33,1} &= E_{11} + E_{12} - E_{22} + E_{23} - E_{31} - E_{33} \leq S_{33,1}^C = 4,
    \label{eq: S_331}
\end{align}
and
\begin{align}
    S_{33,2} = E_{11} &+ E_{12} + E_{13} + E_{21} \nonumber\\
              &- E_{22} - E_{23} - E_{31} + E_{32} - E_{33} \leq S_{33,2}^C = 5.
    \label{eq: S_332}
\end{align}
To the best of our knowledge, the corresponding quantum maxima were previously unknown. The maximal qubit values of these witnesses are derived analytically in Appendix~\ref{app: analytical_Q_bounds}, where we obtain $S_{33,1}^Q = 3\sqrt{3}$ and $S_{33,2}^Q = 6$. Having fixed the witnesses we will use, we next recall the randomness metrics that translate a given violation into certified unpredictability.

Among the various protocols described within the PAM framework, randomness generation protocols are of particular relevance. Their goal is to certify that the randomness extracted from the observed data in a given PAM setup is genuinely quantum. Specifically, the observer must be able to quantify the amount of randomness produced based solely on the empirical distribution $p(b|x,y)$.

The non-trivial nature of this task can be illustrated with a simple example~\cite{Lunghi_SelfTestingQRNG}. Consider a preparation device governed by a hidden variable $\lambda$ that, for each input $x$, sends one of two states: $\rho_x^{\lambda_0} = \ketbra{0}{0}$ or $\rho_x^{\lambda_1} = \ketbra{1}{1}$. Suppose the pre-shared source distributes $\lambda_0$ and $\lambda_1$ with equal probability, $p(\lambda_{0,1}) = \tfrac{1}{2}$, and the measurement device implements  $M_y = \sigma_z$. Then the observed distribution is uniformly random $p(b|x,y) = \tfrac{1}{2}$. However, conditioned on $\lambda$, the outcome is deterministic $p(b=0|x,y,\lambda_0)=1,p(b=1|x,y,\lambda_1)=1$, so randomness is only apparent. In fact, $\lambda$ could be generated by a pseudorandom number generator, and if its algorithm/seed is known to an adversary, the outputs are fully predictable, and the protocol is entirely deterministic.

On the other hand, the same uniform distribution $p(b| x,y)=\tfrac12$ arises if Alice prepares the state $\ketbra{+}{+}$ and Bob measures the spin component $\sigma_z$. In this case, the outcome reflects genuine quantum randomness due to the uncertainty principle. The central challenge, therefore, is to distinguish between these two situations using only the observed statistics. PAM nonclassicality witnesses are designed precisely for this task: they translate a measured violation into a bound on an adversary’s predictive power. Formally, given a witnessed value $W^*$ (i.e., a violation of a classical bound by a specified amount), we upper-bound the best possible single-shot guessing probability over all quantum-realizable preparations and measurements consistent with that violation:
\begin{equation}
    p_\mathrm{guess} := \max_{b,x,y}\, p(b|x,y),
    \label{eq: worst_min_entropy}
\end{equation}
This corresponds to solving the following optimization problem~\cite{li2012semi}:
\begin{align}
     \max_{\{\rho_x\},\{M_{b|y}\}} \:\:\: p_{\mathrm{guess}} &=  \label{eq:max_p_opt} \max_{\{\rho_x\},\;\{M_{b|y}\}} \max_{b,x,y} \; p(b|x,y), \\[4pt]
    \text{s.t.} \:\:
     p(b|x,y) &= \tr\left(\rho_x M_{b|y}\right) \:\:,\:\:\sum_{b,x,y} W_{b|x,y}\, p(b|x,y) = W^*\nonumber.
\end{align}

The guessing probability $p_\mathrm{guess} = \max_{b,x,y} p(b|x,y)$ in Eq.~\eqref{eq: worst_min_entropy} represents the global guessing probability, which is the worst-case over all inputs $x\in \{1..., n_x\}, y\in \{1..., n_y\}$ and outcomes $b$. This metric, while simple, is often too stringent. For instance, if even one pair of settings $(x_d, y_d)$ produces a deterministic output, then $p_\textrm{guess}=1$. This would imply the entire protocol certifies zero randomness, even if other settings $(x', y')$ produce highly random outcomes.

To address this, we can use more fine-grained metrics, where the choice depends on the certification goal. A first approach is for conditional randomness. If the goal is to certify the randomness for the best possible setting (e.g., by post-selecting on reliable rounds), the certifiable randomness can depend on the inputs. For this, we define the conditional guessing probability $p_\mathrm{guess}(x,y) = \max_{b} p(b|x,y)$, the best success probability given $\rho_x$ and $M_y$~\cite{Brown_2020}. A second approach is for average randomness: if the goal is to quantify the average randomness per round (assuming inputs are chosen according to some distribution), a more natural figure of merit is the average guessing probability $\overline{p}_\mathrm{guess}$~\cite{Li2015}, defined as

%The guessing probability $p_\mathrm{guess} = \max_{b,x,y} p(b|x,y)$ is the worst-case over all inputs $x\in \{1..., n_x\}, y\in \{1..., n_y\}$ and outcomes $b$. For any pair $(x_d, y_d)$ that produces a deterministic output, then $p_\textrm{guess}=1$. Yet the protocol may still contain randomness that could be extracted by discarding rounds corresponding to inputs $(x_d, y_d)$. Equivalently, the certifiable randomness can depend on the inputs, so we define the conditional guessing probability $p_\mathrm{guess}(x,y) = \max_{b} p(b|x,y)$, the best success probability given $\rho_x$ and $M_y$~\cite{Brown_2020}. Another natural figure of merit \ABT{Até aqui não ficou muito clara a necessidade de introduzirmos todas essas quantidades. Talvez ficaria mais claro, se explicássemos o que faremos com cada uma delas.} is the average guessing probability $\overline{p}_\mathrm{guess}$~\cite{Li2015}, defined as
\begin{equation}
    \overline{p}_\mathrm{guess} := \sum_{x,y} p(x)p(y) \max_{b} p(b|x,y),
    \label{eq: average_p_guess}
\end{equation}
which quantifies the maximum probability of correctly guessing any outcome $b$, averaged over all experimental realizations, each weighted by $p(x)$ and $p(y)$. As before, to certify randomness, one must optimize over all possible quantum realizations:
\begin{align}
\max_{\{\rho_x\},\,\{M_{b|y}\}} \:\:\:\:\: \overline{p}_\mathrm{guess} &= \label{eq: max_pg_avg}\max_{\{\rho_x\},\,\{M_{b|y}\}} \sum_{x,y} p(x)p(y) \max_{b} p(b|x,y), \\
    \text{s.t.} \quad p(b|x,y) &= \tr(\rho_x M_{b|y})\:\:,\:\:\sum_{b,x,y} W_{b|x,y}\, p(b|x,y) = W^*. \nonumber
\end{align}

A particular case of Eq.~\eqref{eq: average_p_guess} that is worth highlighting---and that will be used throughout this work---is the uniform setting, where $p(x) = 1/n_x$ and $p(y) = 1/n_y$ for all inputs. The corresponding average guessing probability is
\begin{equation}
    \overline{p}^{(u)}_{\mathrm{guess}} := \frac{1}{n_x n_y} \sum_{x,y} \max_b \, p(b|x,y).
    \label{eq: uniform_avg_pguess}
\end{equation}
For instance, in a $(3,2,2,2)$ scenario, $\overline{p}^{(u)}_{\mathrm{guess}} \!=\! \tfrac16 \sum_{x,y} \max_b  p(b|x,y),$
while in a $(3,3,2,2)$ scenario, $\overline{p}^{(u)}_{\mathrm{guess}} \!=\! \tfrac19 \sum_{x,y} \max_b  p(b|x,y)$. 
Naturally, the same optimization problem described in Eq.~\eqref{eq: max_pg_avg} can be formulated for this uniform case, allowing its maximum to be calculated.

Finally, to quantify the amount of certifiable randomness, we use the (conditional) min-entropy, the standard metric in cryptography and randomness certification. It captures the number of secure bits extractable in the worst case. It is calculated from the relevant optimized guessing probability, $p_g^*$, which represents the solution to one of the optimization problems above (e.g., Eq.~\eqref{eq:max_p_opt} or Eq.~\eqref{eq: max_pg_avg}). The min-entropy is defined as:
\begin{equation}
    H_{\infty} = -\log_2 ( p_g^* ).
    \label{eq:min_entropy}
\end{equation}
Here, $p_g^*$ corresponds to the specific certification goal we are interested in. For instance, for global randomness, $p_g^*$ is the optimized global $p_\mathrm{guess}$ from Eq.~\eqref{eq:max_p_opt}. For average randomness, $p_g^*$ is the optimized average $\overline{p}_\mathrm{guess}$ from Eq.~\eqref{eq: max_pg_avg}~\cite{Li2015}. Finally, for conditional randomness, we can define a setting-dependent min-entropy, $H_{\infty}(x,y) = -\log_2(p_\mathrm{guess}^*(x,y))$, where $p_\mathrm{guess}^*(x,y)$ is the optimized conditional guessing probability for a specific pair of inputs $(x,y)$.

Physically, a smaller maximum guessing probability means larger min-entropy and thus more certifiable randomness.

\subsection{Continuous-variable measurement schemes}

In the dimensionally constrained PAM scenarios considered here, one could in principle certify randomness by numerically optimizing over arbitrary states and measurements~\cite{Li2015, chen2021certified, pawlowski2011, li2012semi}. In this work, we restrict to states of the form $\alpha\ket{0}+\beta\ket{1}$ in the Fock basis (first two levels) and to two experimentally standard measurement schemes: homodyne detection and displacement-based photodetection. Homodyne detection measures one quadrature of the electromagnetic field and yields continuous outcomes, which we discretize by binning to obtain dichotomic observables (see, e.g.,~\cite{cavalcanti2011large, chaves2011, Quintino2012}). Displacement-based photodetection produces discrete outcomes (click/no-click), with the displacement setting the effective measurement basis~\cite{banaszek1996, BanaszekWodkiewicz1999, BraskChaves2012,chaves2018causal}. Below we formalize both schemes and the binning used; the resulting dichotomic measurements are then used in our PAM analysis.

\subsubsection{Homodyne detection}
Homodyne detection measures a rotated quadrature of the electromagnetic field by interfering the signal with a strong local oscillator at a 50:50 beam splitter, followed by balanced photodetection of the output modes~\cite{ulfleonhardt_2005_measuring, lvovsky2009}. The dimensionless quadrature operators, $X=(a+a^\dag)/\sqrt{2}$ and $P=-i(a-a^\dag)/\sqrt{2}$, are conjugated variables analogous to the position and momentum operators of a quantum harmonic oscillator that satisfy the canonical commutation relation $[X,P]=i$. Here, $a$ and $a^\dagger$ are the photon annihilation and creation operators of the input mode, respectively. From this point forward, we use natural units $(\hbar=1)$ and model the measurement explicitly. The continuous-spectrum $X$ quadrature operator has eigenstates $\ket{x}$ and eigenvalues $x\in\mathbb{R}$, i.e., $X\ket{x} = x\ket{x}$, and similarly for the $P$ quadrature operator.

More generally, one can define a rotated quadrature, $X(\theta)$, along an arbitrary direction in phase space as $X(\theta)=U(\theta)^\dagger X\,U(\theta)=(a e^{i\theta}+a^\dag e^{i-\theta})/\sqrt{2}$, where $U(\theta)=e^{-i\theta a^\dagger a}$ rotates $X$ by an angle $\theta$. Experimentally, the rotation angle $\theta$ corresponds to the phase difference between the signal and the local oscillator, so that homodyne detection implements the projective measurement $\ket{x_\theta}\bra{x_\theta}=U(\theta)^\dagger\ket{x}\bra{x}U(\theta)$ onto rotated quadrature eigenstates.

As we are interested in a dichotomic measurement, we choose a measurable set $A^+\subset\mathbb{R}$ and its complement $A^-=\mathbb{R}\setminus A^+$. We output $+1$ if the quadrature result lies in $A^+$ and $-1$ otherwise. The corresponding measurement operator is
\begin{equation}
X_{\pm}(\theta) = \Pi_{A^+}(\theta) - \Pi_{A^-}(\theta),
\end{equation}
where each $\Pi_{A^\pm}(\theta)$ projects onto outcomes in $A^\pm$. These projectors are the rotated versions of those associated with the sets $A^\pm$
\begin{equation}
\Pi_{A^\pm}(\theta) = U(\theta)^\dagger \left( \int_{A^\pm}\textrm{d}x \,\ketbra{x}{x} \, \right) U(\theta).
\end{equation}
To evaluate their action on quantum states, we compute the matrix elements of $\Pi_{A^\pm}(\theta)$ in the Fock basis $\{|n\rangle\}_{n=0}^\infty$, where $\hat n\equiv a^\dagger a$ is the number operator and $|n\rangle$ satisfies $\hat n|n\rangle=n|n\rangle$. Using $U(\theta)=e^{-i\theta \hat n}$, we obtain 
\begin{equation}
\langle n | \Pi_{A^\pm}(\theta) | m \rangle = e^{i (n - m) \theta} \int_{A^\pm} \textrm{d}x \, \psi_n^*(x) \psi_m(x), \label{eq: homodyne_projector}
\end{equation}
where $\psi_n(x) = \langle x | n \rangle$ is the wavefunction
\begin{equation}
    \psi_n(x) = \frac{H_n(x)}{\sqrt{2^n n! \sqrt{\pi}}} \exp \left(-\frac{x^2}{2}\right).
\end{equation}
By choosing a rotation angle $\theta$ and a measurable set $A^+\subset\mathbb{R}$, we bin the continuous outcomes into two results. This binning together with $\theta$ completely specifies $\Pi_{A^\pm}(\theta)$ and reduces the problem to evaluating the integral above, thereby enabling different measurement settings in prepare-and-measure scenarios.

The primary advantage of employing homodyne detection in tests of Bell-like inequalities lies in its ability to achieve detection efficiencies approaching unity. Consequently, such tests are not subject to the usual detection-efficiency loophole, which remains one of the major challenges for implementations based on single-photon detection~\cite{hensen2015loophole,Giustina2015,loop1}. Homodyne schemes have previously been investigated in the context of Bell inequality violations~\cite{cavalcanti2011large, chaves2011, Quintino2012}, yet no violation has been observed relying solely on homodyne measurements. To the best of our knowledge, homodyne detection has not yet been explored within the PAM scenario, in relation to the violation of dimension witness inequalities. As we show below, it is indeed possible to achieve violations of dimension witness inequalities using only homodyne detection. A key advantage of our approach is that it enables a simpler implementation with near-unit detection efficiency.

\subsubsection{Displacement-based photodetection}

Displacement-based photodetection is an experimentally accessible scheme that probes quantum states by combining a phase-space displacement with binary photodetection. Experimentally, the signal is interfered with a highly excited coherent state (local oscillator) on a highly transmissive beam splitter, followed by an on/off photon detector~\cite{banaszek1996,BanaszekWodkiewicz1999,BraskChaves2012,chaves2018causal}.

The measurement consists of first applying the displacement operator $D(\alpha)=e^{\alpha a^\dagger-\alpha^* a}$, which shifts the state in phase space by the complex amplitude $\alpha$. Subsequently, a single-photon detector distinguishes between no-click and click events. Assigning $+1$ to no-click and $-1$ to click, we define the measurement operator $M_D(\alpha)=P_{+}(\alpha)-P_{-}(\alpha)$, where $P_{+}(\alpha)=D(\alpha)\ketbra{0}D^\dagger(\alpha)=\ketbra{\alpha}$ and $P_{-}(\alpha)=\iden-P_{+}(\alpha)$. Here, $|\alpha\rangle=D(\alpha)\ket{0}$ is the coherent state with complex amplitude $\alpha$ and $\iden$ is the identity operator. As a consequence, $M_D(\alpha)=2\ketbra{\alpha}-\iden$.

The matrix elements of these operators in the Fock basis are
\begin{equation}
\langle n | P_{+}(\alpha) | m \rangle = e^{-|\alpha|^2} \frac{\alpha^{m} (\alpha^*)^{n}}{\sqrt{n! m!}}, \label{eq: displacement_projector}
\end{equation}
and those of $P_{-}(\alpha)$ follow from completeness. 

By varying the displacement parameter $\alpha = re^ {i\varphi}$, different measurement settings for PAM scenarios can be implemented. In the subspace $\{|0\rangle,|1\rangle\}$, the matrix representing $M_D$ is
\begin{equation}
    M_D(r,\varphi)= 
    2e^{-r^2}\begin{pmatrix}
    1 & r e^{-i \varphi} \\ r e^{i\varphi} & r^2
\end{pmatrix} - \iden_2,
\label{eq:matrix_displacement}
\end{equation}
where $\iden_2$ represents the two-dimensional identity matrix\footnote{To avoid ambiguity, we write $\iden_2$ for the identity on the qubit space (truncated subspace). This distinguishes it from $\iden$, the identity on the Fock space.}. In practice, finite detection efficiency and dark counts make the elements $P_{\pm}(\alpha)$ non-projective POVM elements. The ideal expressions above are recovered in the unit-efficiency, zero–dark-count limit. For completeness, one may model imperfections by replacing $P_{+}(\alpha)$ with the no-click POVM element of an on/off detector after displacement, and set $P_{-}(\alpha)=\iden-P_{+}(\alpha)$. The resulting $M_D(\alpha)$ can then be evaluated in any truncated Fock subspace as above.

\section{Results \label{Sec:results}}

\subsection{Optimal witnesses values with homodyne and displacement-based detection} \label{Sec:sub-homo-dis-detection}
\label{sec: resultsA}
To evaluate the optimal value of a witness attainable with homodyne and displacement measurements, we formulate an optimization problem over both the prepared states and the measurement settings. Formally, this can be written as
\begin{align}
    \max_{\{\rho_x\}, \{M_{b|y}\}} \:\: W  = 
&\max_{\{\rho_x\}, \{M_{b|y}\}} \sum_{b,x,y} W_{b|x,y}\,p(b|x,y) \nonumber
    \\ &\quad \:\text{s.t.} \quad p(b|x,y) = \tr( \rho_x M_{b|y}), \label{eq: maxW}
\end{align}
where $W$ is the witness value with fixed real coefficients and the POVM elements $M_{b|y}$ are chosen from the measurement families in Eqs.~\eqref{eq: homodyne_projector} and \eqref{eq: displacement_projector} (homodyne and displacement, respectively).

\begin{table}[t!]
\centering
\caption{Optimal numerical values of the dimension witnesses $S_3$ and $S_4$ for different measurement setups. HH: homodyne for $y=1,2$; DD: displacement for $y=1,2$; HD: homodyne for $y=1$ and displacement for $y=2$; DH: displacement for $y=1$ and homodyne for $y=2$.}
\label{tab: witness_values}
\small
\setlength{\tabcolsep}{6pt}
\renewcommand{\arraystretch}{1.2}
\begin{tabular}{c|cc}
\hline\hline
Measurement schemes & $S_3$ & $S_4$ \\
\hline
HH setup & 3.112 & 4.512 \\
DD setup & 3.783 & 5.592 \\
HD setup & 3.418 & 5.116 \\
DH setup & 3.558 & 5.116 \\
\hline\hline
\end{tabular}
\end{table}

Since we restrict to the two–dimensional subspace spanned by $\{\ket{0}, \ket{1}\}$, we may, without loss of generality, take the preparations to be pure and parameterize them as $\rho_x = \ketbra{\varphi(x)}$, where 
\begin{align}
\label{eq: rho_bloch_parameterized}
\ket{\varphi(x)} = \cos\qty(\frac{\alpha_x}{2})\ket{0} + e^{i\eta_x}\sin\qty(\frac{\alpha_x}{2})\ket{1}
\end{align}
with $x \in \{1,\dots,n_x\}$, $0 \leq \alpha_x \leq \pi$, and $0 \leq \eta_x < 2\pi$. 
% For each choice $y \in \{1, \dots, n_y\}$, the measurement is a POVM $\{M_{b|y}\}_{b}$, selected from homodyne or displacement. In the $\{\ket{0},\ket{1}\}$ subspace these POVM elements are $2 \times 2$ matrices whose entries are given explicitly by Eqs.~\eqref{eq: homodyne_projector} and \eqref{eq: displacement_projector}.
Reliable schemes for generating such states rely on the quantum scissors protocol~\cite{pegg1998optical}, which has already been demonstrated experimentally~\cite{babichev2003quantum}.

To complete the specification of the optimization, we parametrize the measurements and restrict their domains. For homodyne measurements, we bin outcomes with $A^+=[a,b]$ and restrict the quadrature angle to $0\le \theta\le \pi$, treating $a$, $b$, and $\theta$ as optimization variables. For numerical evaluation of Eq.~\eqref{eq: homodyne_projector}, we bound $a,b$ to $[-5,5]$. Because the Hermite–Gaussian quadrature wavefunctions decay exponentially, contributions outside this window are negligible and effectively reproduce the full-line integral. For displacement measurements, we write $\alpha=r e^{i\phi}$ with $0\le r\le 1$ and $0\le \phi<2\pi$, and optimize over $r$ and $\phi$.

Within this optimization framework, we determine the optimal values of the dimension witnesses for the different scenarios considered by employing nonlinear optimization.
\begin{figure*}[t]
    \centering
    \includegraphics{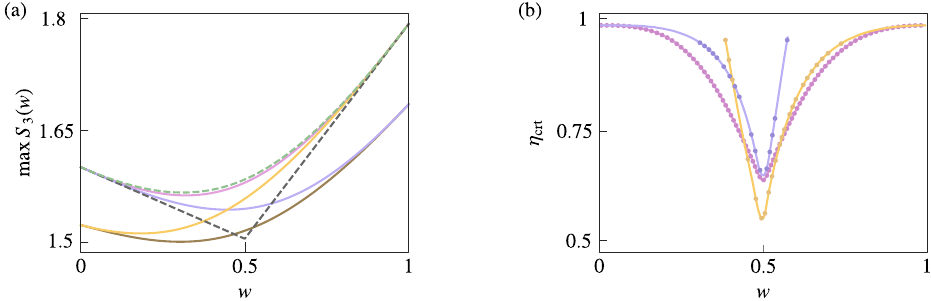}
    \caption{\textbf{Tilt dependence of violation and efficiency}. (a) Optimal values of $S_3(w)$ as a function of the tilt parameter $w$ for different measurement setups. The results for both qubit (dashed green) and bit (dashed gray) systems using general measurements are also included for comparison. (b) Critical detection efficiency $\eta_{\mathrm{crt}}$ required to violate the tilted inequality $S_3(w)$ in the $(3,2,2,2)$ scenario, shown as a function of $w$. The color code denotes the measurement setups as follows: HH (brown), DD (magenta), HD (violet), DH (yellow). The measurement labels follow the same notation as in Table~\ref{tab: witness_values}.}
    \label{fig: s3_tilted}
\end{figure*}

To exemplify this approach, we analyze the $(3,2,2,2)$ and $(4,2,2,2)$ scenarios, with witnesses $S_3$ and $S_4$ [Eqs.~\eqref{eq:dw_ineqS3} and~\eqref{eq: s4}], under different measurement schemes. The results are summarized in Table~\ref{tab: witness_values} for full homodyne (HH), full displacement (DD), and hybrid schemes (HD and DH). In the hybrid cases, each measurement setting $y$ is assigned a different measurement type ---for example, HD denotes homodyne for $y=1$ and displacement for $y=2$ (resp. $y=0$ and $y=1$ if zero-indexed), while DH denotes the opposite assignment. All schemes considered achieve violations of the corresponding dimension-witness inequalities under ideal conditions. The optimal states and measurement parameters are listed in Appendix~\ref{app: opt_params}. We emphasize that the displacement-based scheme achieves violations of the $S_3$ and $S_4$ inequalities that approach the maximum values attainable with optimal general projective measurements. An even more notable feature is that, using only homodyne detection, we still observe a significant violation of these inequalities, something not achievable in standard Bell tests~\cite{cavalcanti2011large,Quintino2012,chaves2011}. This demonstrates that loophole-free violations of dimension-witness inequalities can be realized in a much simpler experimental setup.

We also consider the tilted $S_3(w)$ [Eq.~\eqref{eq: s3_tilted}] in the $(3,2,2,2)$ scenario. For each $w$ in a discrete partition of $[0,1]$, we solve the optimization in Eq.~\eqref{eq: maxW} for all measurement schemes. Figure~\hyperref[fig: s3_tilted]{3(a)} summarizes the results and shows that the largest quantum–classical gap occurs near $w=0.5$ across all measurement types. Once again, the displacement scheme attains violations very close to the maximum values achievable with optimal general projective measurements. For the tilted inequality, however, the homodyne scheme exhibits a violation only within a narrow range of the tilting parameter.
\begin{figure*}
    \centering
    \includegraphics{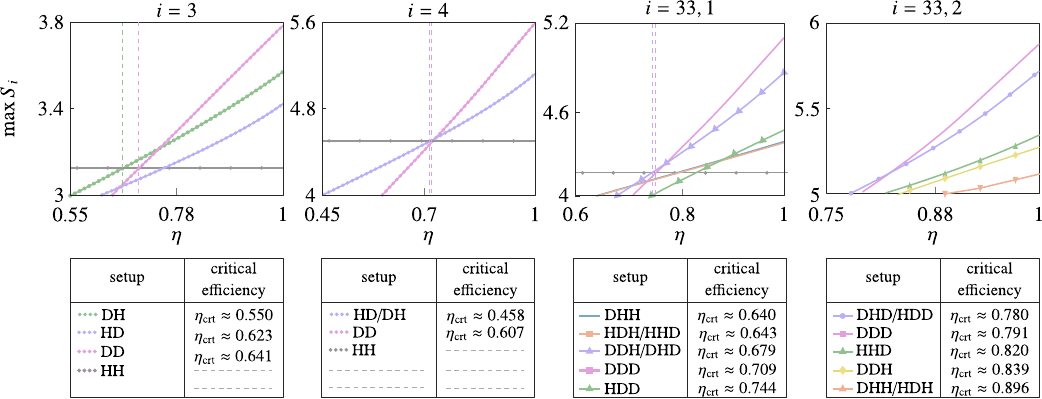}
    \caption{\textbf{Witness violation versus detection efficiency}. Maximum attainable value of the PAM witnesses (left to right) $i \in {S_3, S_4, S_{33,1}, S_{33,2}}$ as a function of the displacement detection efficiency $\eta$. Losses are modeled on the displacement stage while homodyne is treated as lossless. Each curve corresponds to a measurement setup (DD/HD/DH for two settings; DDD/DDH/$\ldots$ for three). A violation occurs above the classical bound, and the intersection defines the critical efficiency $\eta_{\rm crit}$. Dashed lines indicate the values of $\eta_{>{\rm HH}}$ at which hybrid or displacement configurations overtake the constant HH curve, marking the efficiency thresholds beyond which they outperform full homodyne detection.}
    \label{F:max-attainable}
\end{figure*}
We further extend our analysis to the $(3,3,2,2)$ scenario, which features the witness $S_3$ together with two additional inequalities, $S_{33,1}$ and $S_{33,2}$ [Eqs.~\eqref{eq: S_331} and~\eqref{eq: S_332}]. With three measurement settings, there are multiple configurations: each $y\in\{1,2,3\}$ can independently be homodyne (H) or displacement (D), e.g., HDD, HDH, DHD, etc. (see Table~\ref{tab:configs_332}). We omit $S_3$ from the table since it involves only two measurements and thus coincides with the $(3,2,2,2)$ case; its maximal values are therefore unchanged. In this three-setting scenario, the fully homodyne configuration (HHH) is the only non-violating setup for $S_{33,2}$. All other configurations yield violations of both $S_{33,1}$ and $S_{33,2}$. The optimal state and measurement parameters for representative configurations (DDD and HHH) are listed in Appendix~\ref{app: opt_params}.
\begin{table}[t]
\centering
\caption{Optimal values of the dimension witnesses $S_{33,1}$ and $S_{33,2}$ for different measurement configurations in the $(3,3,2,2)$ scenario. In each configuration, the letters indicate the measurement type, homodyne (H) or displacement (D), for each setting $y \in \{1,2,3\}$, listed in order from $y=1$ to $y=3$.}
\label{tab:configs_332}
\small
\setlength{\tabcolsep}{6pt}
\renewcommand{\arraystretch}{1.2}
\begin{tabular}{c|cc}
\hline\hline
Measurement schemes & $S_{33,1}$ & $S_{33,2}$ \\
\hline
DDD setup & 5.0899 & 5.8251 \\
DDH setup & 4.8497 & 5.2533 \\
DHD setup & 4.8497 & 5.6749 \\
DHH setup & 4.3740 & 5.1050 \\
HDD setup & 4.4531 & 5.6749 \\
HDH setup & 4.3659 & 5.1049 \\
HHD setup & 4.3659 & 5.3210 \\
HHH setup & 4.1459 & 4.9076 \\
\hline\hline
\end{tabular}
\end{table}

\subsection{Detection efficiencies for displacement and combined measurements} \label{Sec:detection-eff-disp-cm}

We next model detection inefficiencies via an amplitude-damping channel that captures photon losses~\cite{nielsen2010,2025detection}. The system couples to a vacuum environment, inducing the decay $\ket{1}\to\ket{0}$ with probability $1-\eta$, where $\eta\in[0,1]$ denotes the detection efficiency. The CPTP channel can be written as 
\begin{equation}
    \Lambda_\eta(\rho) = E_0\rho E_0^\dagger + E_1\rho E_1^\dagger,
    \label{eq: amplitude_damping}
\end{equation}
with the Kraus operators $E_0 \!=\! \ketbra{0} + \sqrt{\eta}\ketbra{1}$ and $E_1 \!=\! \sqrt{1-\eta}\,\ket{0}\bra{1}$ suming up to the identity, i.e., $E_0^\dagger E_0 + E_1^\dagger E_1 = \iden_2$. Including losses in Eq.~\eqref{Eq:PAM-born-rule} yields $p_\eta(b|x,y) = \tr[\Lambda_\eta(\rho_x) M_{b|y}]$. Equivalently, by introducing the adjoint channel $\Lambda_\eta^\dagger$ defined by $\tr[\Lambda_\eta(\rho)M]=\mathrm{Tr}[\rho\Lambda_\eta^\dagger(M)]$, losses can be viewed as acting either on the states or on the measurement operators, without changing the resulting statistics. In our numerical implementation, we apply $\Lambda_\eta$ to the states for convenience, interpreting $\eta$ as the detection efficiency.

We are interested in determining the maximum value attainable by a PAM witness when the detection efficiency is fixed at $\eta=\eta^{*}$. This is formulated analogously to Eq.~\eqref{eq: maxW}, but with the Born rule modified to include losses:
\begin{align}
\label{eq: maxW_lossy}
    \max_{\{\rho_x\}, \{M_{b|y}\}} W  = &\max_{\{\rho_x\}, \{M_{b|y}\}} \sum_{b,x,y} W_{b|x,y}\,p_\eta(b|x,y) \notag
    \\ &\quad\:\:\text{s.t.} \quad p_\eta(b|x,y) = \mathrm{Tr}[\Lambda_\eta(\rho_x) M_{b|y}], \notag
    \\ &\hspace{1.4cm}\eta = \eta^* \in [0,1],
\end{align}
where $\Lambda_\eta(\rho_x)$ is given by Eq.~\eqref{eq: amplitude_damping} and the states are parameterized as in Eq.~\eqref{eq: rho_bloch_parameterized}. Measurements are chosen from the homodyne and displacement families of Eqs.~\eqref{eq: homodyne_projector} and \eqref{eq: displacement_projector}.

Inefficiencies are modeled only for displacement and combined displacement–homodyne scenarios, since homodyne detection is typically treated as ideal owing to its near-unit quantum efficiency~\cite{lvovsky2009,Serikawa2018,Wang2024}. Consequently, Eq.~\eqref{eq: maxW_lossy} is applied only to the settings $y$ corresponding to displacement-based measurements, while homodyne settings use the lossless formulation in Eq.~\eqref{eq: maxW}.

We solve the optimization accounting for detection losses for the PAM witnesses in the $(3,2,2,2)$, $(4,2,2,2)$, and $(3,3,2,2)$ scenarios by instantiating $W$ with Eqs.~\eqref{eq:dw_ineqS3}, \eqref{eq: s3_tilted}, \eqref{eq: s4}, \eqref{eq: S_331}, and \eqref{eq: S_332}. For each fixed efficiency $\eta=\eta^*\in[0,1]$, we compute the maximal attainable value of each witness. The \emph{critical efficiency} $\eta_{\mathrm{crt}}$ is the smallest $\eta$ for which a violation is still possible; for $\eta<\eta_{\mathrm{crt}}$ no violation can occur.

For the $(3,2,2,2)$ scenario, the tilted and standard $S_3$ results appear in Fig.~\hyperref[fig: s3_tilted]{3(b)} and Fig.~\ref{F:max-attainable}, respectively. Figure~\ref{F:max-attainable} also reports results for the $(4,2,2,2)$ and $(3,3,2,2)$ cases for $S_4$, $S_{33,1}$, and $S_{33,2}$. The $S_3$ inequality is omitted for $(3,3,2,2)$ since it involves only two measurements and thus matches the $(3,2,2,2)$ results. In all cases, the attainable witness value decreases as the detection efficiency is reduced. When considering displacement measurements as well, the hybrid approach exhibits the greatest robustness against detection inefficiencies. The most resilient case we could identify corresponds to the $S_4$ inequality, for which the critical detection efficiency of the photon detector used in the displacement measurement is $\eta_{crt} \approx 0.458$.

Since homodyne measurements can be implemented with near-unit efficiency, they are free from the detection loophole. However, this advantage comes at the cost of yielding smaller violations compared to displacement-based schemes. Beyond a critical efficiency threshold, denoted by $\eta_{>{\rm HH}}$, hybrid or displacement setups outperform full homodyne detection and thus become the preferred choice, as illustrated in Fig.~\ref{F:max-attainable}.

\subsection{Min-entropy bounds}
\subsubsection{Analytical bound for the (3,2,2,2) scenario}
Following~\cite{Li2015,Lunghi_SelfTestingQRNG}, we derive an analytical upper bound on the uniform average guessing probability defined in Eq.~\eqref{eq: uniform_avg_pguess}, focusing on the $(3,2,2,2)$ scenario ($n_x=3$, $n_y=2$). In particular, we evaluate this bound for the tilted inequality $S_3(w)$ introduced in Eq.~\eqref{eq: s3_tilted}. The full derivation is given in Appendix~\ref{app: analytical_upper_bound}. The resulting expression reads
\begin{equation}
    \overline{p}_{\mathrm{guess}}^{(u)} \leq F[S_3(w)],
\end{equation}
where $\overline{p}_{\mathrm{guess}}^{(u)} = 1/6\sum_{x,y} \max_b p(b|x,y)$ for this scenario, and
\begin{align}
    F[S_3(w)] = \frac{1}{2} + \frac{1}{2}\sqrt{\frac{1}{2} + \frac{A^2 - \big[(S_3(w)-w)^2/2 - A\big]^2}{2B^2}},
\end{align}
with $A = w^2 + (1-w)^2$ and $B = 2w(1-w)$.

This upper bound also gives a lower bound on the generated randomness, since the min-entropy in Eq.~\eqref{eq:min_entropy} obeys
\begin{equation}
    H_\infty\big(F[S_3(w)]\big) \leq H_\infty\big(\overline{p}_{\mathrm{guess}}^{(u)} \big).
    \label{eq: hmin_lower_bound}
\end{equation}
Thus, $F[S_3(w)]$ provides a computable analytical estimate of the randomness guaranteed in any implementation, independent of the numerical optimization. Fig.~\ref{fig:Hmin_s3_tilted} compares the analytical bound with the min-entropy obtained from the numerical optimization discussed below, plotted as a function of the normalized tilted witness for several values of \(w\). To compare different values of $w$ on the same scale, we use the normalized quantity $\frac{S_3(w)-S_3^C(w)}{S^Q_3(w)-S_3^C(w)}$ in the horizontal axis, so that $0$ corresponds to the classical bound and $1$ to the quantum maximum. In this way, the horizontal axis directly reflects the relative strength of the violation.

\subsubsection{Numerical bounds for the (3,2,2,2) and (3,3,2,2) scenarios}
In this section we compute the optimization in Eq.~\eqref{eq: max_pg_avg} for the $(3,2,2,2)$ and $(3,3,2,2)$ scenarios, following the approach used for $(4,2,2,2)$ in Ref.~\cite{Li2015}.

\begin{figure}[t!]
    \centering
    \includegraphics{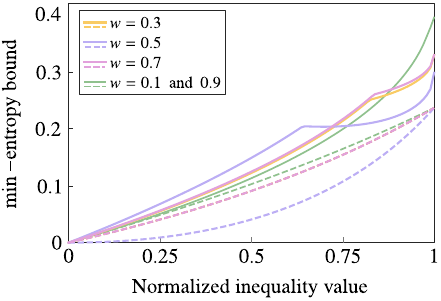}
    \caption{
    \textbf{Min-entropy versus witness violation}. Min-entropy $H_{\infty}$ of the uniform average guessing probability 
    $\overline{p}^{(u)}_{\mathrm{guess}} = 1/6 \sum_{x,y} \max_b p(b|x,y)$ as a function of the normalized value of the tilted witness $S_3(w)$ for several values of \(w\). The normalization is such that $0$ corresponds to the classical bound and $1$ to the quantum maximal violation. The solid lines show the min-entropy obtained from the optimized uniform average guessing probability, while the dashed lines correspond to the analytical lower bound given in Eq.~\eqref{eq: hmin_lower_bound}. The maximal min-entropies for the rightmost points of each curve follows. For the optimized values, $H_{\infty}^{\max} \approx 0.382$ for $w = 0.10$ and $w = 0.90$; 
    $H_{\infty}^{\max} \approx 0.318$ for $w = 0.30$ and $w = 0.70$; 
    $H_{\infty}^{\max} \approx 0.288$ for $w = 0.50$.
    For the analytical lower bound, $H_{\infty}^{\max} \approx 0.228$ for all $w$ values.}
    \label{fig:Hmin_s3_tilted}
\end{figure}

To solve the problem, we first parametrize states and measurements in the two-dimensional Hilbert space. States are written as in Eq.~\eqref{eq: rho_bloch_parameterized}; this suffices because the witnesses depend linearly on the probabilities, so optimal preparations can be taken pure. Since the measurements are dichotomic, they can, without loss of generality, be taken as rank-1 projectors~\cite{Masanes2005Extremal}. We parameterize them as:
\begin{align}
M_{0|1} &=
\begin{pmatrix}
1 & 0 \\
0 & 0
\end{pmatrix}\nonumber, \\
M_{0|y} &=
\begin{pmatrix}
\cos^2\!\qty(\tfrac{\beta_y}{2}) & \tfrac{1}{2} e^{-i\gamma_y} \sin(\beta_y) \\
\tfrac{1}{2} e^{i\gamma_y} \sin(\beta_y) & \sin^2\!\qty(\tfrac{\beta_y}{2})
\end{pmatrix},
\end{align}
where $y>1$, $0\le \beta_y\le \pi$, and $0\le \gamma_y<2\pi$. The first measurement, $M_{0|1}=\ket{0}\bra{0}$, is fixed along the $z$-axis of the Bloch sphere to remove the global rotational freedom: any simultaneous unitary rotation of all states and measurements leaves the probabilities invariant.

Using nonlinear optimization, we solve Eq.~\eqref{eq: max_pg_avg} for the witnesses $S_3(w)$, $S_{33,1}$, and $S_{33,2}$, with the uniform average guessing probability defined in Eq.~\eqref{eq: uniform_avg_pguess} --namely, $\tfrac{1}{6}\sum_{x,y}\max_b p(b|x,y)$ for the $(3,2,2,2)$ scenario and $\tfrac{1}{9}\sum_{x,y}\max_b p(b|x,y)$ for the $(3,3,2,2)$ scenario.

\begin{figure}[t!]
    \centering
    \includegraphics{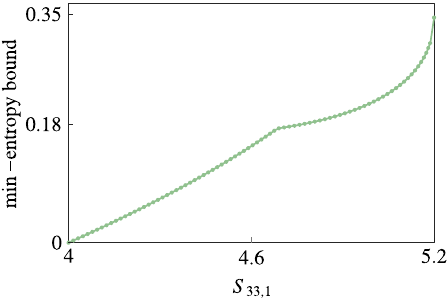}
    \caption{\textbf{Min-entropy versus $S_{33,1}$ violation}. Min-entropy $H_{\infty}$ of the uniform average guessing probability $1/9 \sum_{x,y} \max_b p(b|x,y)$ as a function of the witness $S_{33,1}$. The last point corresponds to the maximal achievable qubit violation, with $H_{\infty}^{\max} = 0.342$.}
    \label{fig:Hmin_S331}
\end{figure}

The optimization is carried out over all valid pure-state preparations and rank-1 projective measurements described above. The resulting optimal uniform average guessing probabilities are converted into min-entropy via Eq.~\eqref{eq:min_entropy}. Figure~\ref{fig:Hmin_s3_tilted} reports the min-entropy for the normalized values of the tilted witness $S_3(w)$ across several values of $w$, together with the analytical lower bound from Eq.~\eqref{eq: hmin_lower_bound}. As can be seen, the tilting parameter enables a significant increase in the amount of randomness, that goes from $H_{\infty}^{\max} \approx 0.288$ for $w = 0.50$ (corresponding to the $S_3$ inequality) to $H_{\infty}^{\max} \approx 0.382$ for $w = 0.10$ and $w = 0.90$. For the three–$y$-input witnesses $S_{33,1}$ and $S_{33,2}$, the maximal min-entropies at the respective quantum maxima are
\begin{equation}
\begin{aligned}
H_{\infty}^{\max}(S_{33,1}) &= 0.342, \\
H_{\infty}^{\max}(S_{33,2}) &= 0.263,
\end{aligned}
\end{equation}
and the complete min-entropy profile for $S_{33,1}$ as a function of the observed violation is shown in Figure~\ref{fig:Hmin_S331}.

For the $(4,2,2,2)$ scenario, the behavior of the min-entropy curve has already been investigated in~\cite{Li2011,li2012semi}, where the optimal guessing probability of Eq.~\eqref{eq: worst_min_entropy} is computed as a function of the $S_4$ violation. This yields a maximal certifiable randomness of \(H_{\infty}^{\max}(S_4) = 0.2284\). The same inequality was later analyzed in~\cite{Li2015} under the uniform guessing probability assumption, providing the full min-entropy curve for $S_4$ with the same maximal min-entropy as in the case of Eq.~\eqref{eq: worst_min_entropy}. Furthermore,~\cite{li2012semi} extends the analysis to the broader class of $(n^2,n,2,2)$ scenarios, closely connected to \(n \to 1\) quantum random access codes~\cite{ambainis2009quantumrandomaccesscodes}. The best scenario found in~\cite{li2012semi}, with \(H_{\infty}^{\max}= 0.3425\), was in a \(3 \to 1\) quantum random access code, that is, a PAM scenario with eight preparations and three different measurement setups. Therefore, our results demonstrate that a simpler scheme, requiring significantly fewer state preparations and measurement settings, can certify more randomness than the previously best-known approach. 

Overall, these results show that any violation of the classical bound directly translates into a positive amount of certifiable randomness. In particular, since the min-entropy increases monotonically with the degree of violation~\cite{Li2015}, every nonclassical value of a PAM witness quantifies the strength of the underlying quantum randomness generation.

As our results prove, considering only facet inequalities, the $S_{33,1}$ inequality provides the best possible amount of randomness. However, tilting the inequality $S_3$ improves the amount of certifiable randomness, something similar to what happens in Bell scenarios \cite{acin2012randomness}.

\subsection{Free reference frames}

Aligned reference frames between the parties are usually assumed to be necessary in order to observe violations of Bell inequalities. However, this requirement was later called into question~\cite{shadbolt2012guaranteed}, where it was shown that neither alignment nor device calibration is needed to obtain near-certain Bell violations. In this section, we pursue an analogous result in PAM scenarios, showing that shared reference frames are likewise unnecessary to violate dimension witnesses.
\begin{figure}
    \centering
    \includegraphics{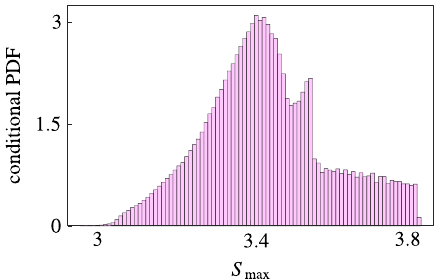}
    \caption{\textbf{Witness violation distribution}. Histogram of the maximal value $S_{\max}$ of the $(3,2,2,2)$ dimension-witness $S_3$ obtained over all relabellings, when Bob’s measurements are uniformly rotated (free reference frame). We plot the conditional PDF of $S_{\max}$ given $S_{\max}>3$. The distribution is normalized over the violating runs only with bin width $\Delta S=0.01$. Fewer than $0.1\%$ of random rotations fail to violate, while the distribution peaks near $S_{\max}\approx 3.4$ with a secondary peak around $3.55$.}
    \label{fig:pdf_free-ref-frame_proj}
\end{figure}

\begin{figure*}
    \centering
    \includegraphics{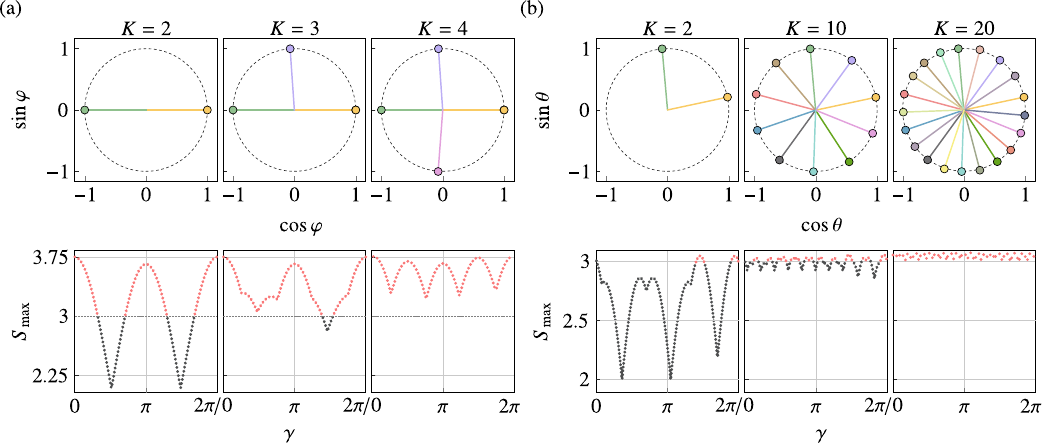}
    \caption{\textbf{Witness robustness to phase misalignment.} \textbf{(a) Displacement} and \textbf{(b) Homodyne} measurements with free reference frames. Top panels: measurement configurations—phase-space locations $(r_y,\varphi_y)$ for displacement, angular positions $\theta_y$ for homodyne. Two optimal settings remain fixed while additional measurements ($K$ total) are added. Bottom panels: maximal violation $S_{\max}(\gamma)$ versus Bob's phase offset $\gamma$. Red markers: $S_{\max}>3$; black: $S_{\max}\le3$; dashed line: classical bound $S=3$.}
    \label{F:displacement-free-ref-frame}
\end{figure*}
We first allow Bob to perform generic projective measurements, and subsequently restrict this to displacement and homodyne measurements. All analysis in this section concerns the $(3,2,2,2)$ prepare-and-measure (PAM) scenario, focusing specifically on the inequality $S_3$. For notational simplicity, we will remove the index 3 from $S_3$ (referring to it simply as $S$). Note, however, that a similar analysis can be performed for all other scenarios.

Let us consider projective measurements. We aim to test the violation of the dimension witness given in Eq.~\eqref{eq:dw_ineqS3}. Let $\vec{\sigma} = (\sigma_x,\sigma_y,\sigma_z)$ denote the Pauli matrices. By writing the states and measurements in the Bloch form $\rho_x=\tfrac{1}{2}\bigl(\iden_2+\vec{a}_x\cdot\vec{\sigma}\bigr)$, $M_{0|y}=\tfrac{1}{2}\bigl(\iden_2+\vec{b}_y\cdot\vec{\sigma}\bigr)$, and $M_{1|y}=\iden_2-M_{0|y}$, the correlators $E_{xy}$ can be expressed as $E_{x,y} = - \vec{a}_x \cdot \vec{b}_y$.

As shown in Ref.~\cite{Gallego2010}, an optimal quantum strategy for violating the $(3,2,2,2)$ dimension witness [Eq.~\eqref{eq:dw_ineqS3}] uses
\begin{align}
&\vec a_1 = (1,0,0),\quad \vec a_2 = (0,1,0),\quad \vec a_3 = \tfrac{1}{\sqrt{2}}(-1,-1,0); \\
&\vec b_1 = \tfrac{1}{\sqrt{2}}(1,-1,0),\quad \vec b_2 = \tfrac{1}{\sqrt{2}}(1,1,0).
\end{align}
We extend this set by adding $\vec a_4=\vec b_3=(0,0,1)$ to the state and measurement catalogs, enabling tests over all relevant relabelings of Eq.~\eqref{eq:dw_ineqS3}. The procedure is:

\begin{enumerate}
  \item Select settings: Alice chooses three preparations from $\{\vec a_1,\vec a_2,\vec a_3,\vec a_4\}$ and Bob chooses two measurements from $\{\vec b_1,\vec b_2,\vec b_3\}$.
  \item Model the absence of a shared reference frame by rotating Bob’s system: for each Euler rotation $R(\alpha,\beta,\gamma)$, replace $\vec b_y \mapsto R\vec b_y$.
  \item Compute the correlators $E_{x,y}=-\vec a_x\cdot R\,\vec b_y$. Enumerate all relabelings consisting of (i) ordered triples $(x,x',x'')$ of pairwise distinct preparations from ${1,2,3,4}$, (ii) ordered pairs $(y,y')$ with $y\neq y'$ from ${1,2,3}$, and (iii) all outcome flips. For each relabeling, evaluate 
    \begin{equation}
      S = E_{x,y} + E_{x,y'} + E_{x',y} - E_{x',y'} - E_{x'',y}.
      \label{eq:generic322}
    \end{equation}
  \item For each rotation, record the maximal witness value over all relabelings, $S_{\max}=\max_{\text{relabel}} S$. Repeat over the rotation ensemble to obtain the set ${S_{\max,j}}$, where $j$ indexes the rotations.
\end{enumerate}

To highlight the distribution shape only for rotations that yield significant violations, we define the conditional probability density function (PDF) on the subset $S_{\max,j}>3$. If $N$ is the number of rotations with $S_{\max}>3$ and $\mathrm{count}_k^{(>3)}$ is the number of occurrences with $S_{\max}=s_k$ within a bin of width $\Delta S$, then the conditional PDF is
\begin{equation}
    \mathrm{PDF}^{(\mathrm{cond})}(s_k) \;=\; \frac{\mathrm{count}_k^{(>3)}}{N\,\Delta S},
\end{equation}
where, by normalization, $\sum_k \mathrm{PDF}^{(\mathrm{cond})}(s_k)\Delta S = 1$ when the sum is taken only over bins with $s_k>3$. Fig.~\ref{fig:pdf_free-ref-frame_proj} illustrates the conditional probability density function in this scenario.

Less than $0.1\%$ of all rotations yielded $S_{\max}\le 3$, showing that quantum violations persist for nearly all reference-frame misalignments. As seen in Fig.~\ref{fig:pdf_free-ref-frame_proj}, the distribution peaks near $S_{\max}\approx3.4$, with a secondary peak around $3.55$. Quantitatively, about $42.05\%$ of rotations give $S_{\max}\in[3.0,3.4)$, $41.75\%$ fall in $[3.4,3.6]$, and $16.20\%$ produce stronger violations in $(3.6,1+2\sqrt{2}]$. The highest violations offer the largest margin above the classical bound, providing optimal conditions for certification tasks, while the more moderate violations ---though still quantum--- may benefit from refined experimental control to ensure robust statistical confidence.

For homodyne and displacement measurements, we employ a similar methodology. In both instances, we model each rotation following the approach detailed in Ref.~\cite{brask2013testing}. Crucially, there is no requirement to implement full Euler rotations because, as shown in Eqs.~\eqref{eq: homodyne_projector} and~\eqref{eq:matrix_displacement}, the continuous variable measurements are parameterized by a single angle. The procedure is:

\begin{enumerate}
    \item Use the parameters that give the optimal preparations $(\rho_1,\rho_2,\rho_3)$ and measurements $(M_1,M_2)$.
    \item Add a third measurement to enable testing different relabelings:
\begin{enumerate}
    \item Displacement: given the optimal $M_D(r_1,\varphi_1)$ and $M_D(r_2,\varphi_2)$, choose $(r_3,\varphi_3)$ to maximize the witness violation after relabelings.
\item Homodyne: since there is no radial parameter, set $\theta_3$ at the maximal angular separation from $\theta_1$ and $\theta_2$.
\end{enumerate}
    \item Model Bob’s reference-frame misalignment by adding a phase $\gamma$ to his measurements: $M_D(r,\varphi)\mapsto M_D(r,\varphi+\gamma)$ for displacement and $X(\theta)\mapsto X(\theta+\gamma)$ for homodyne.
    \item Test all relabelings: compute $E_{x,y}=2p(0|x,y)-1$ and evaluate Eq.~\eqref{eq:generic322}.
    \item For each $\gamma$, record the maximal witness value over all relabelings, $S_{\max}(\gamma)=\max_{\text{relabel}} S$. Sweeping $\gamma$ yields the curve $\gamma\mapsto S_{\max}(\gamma)$.
\end{enumerate}
We then increased the number of available homodyne and displacement settings until we reached a regime where, for some relabeling choice, the witness was violated for \emph{every} rotation. Let $K$ denote the total number of measurements available to Bob, from which he must select two to perform the experiment.

These results, shown in Fig.~\ref{F:displacement-free-ref-frame}, demonstrate that even if with a given number $K$ of measurement settings one does not always obtain a violation, increasing the pool of available measurements eventually leads to a regime in which there always exists a pair that violates the witness, irrespective of the reference-frame rotation. Since displacement measurements  achieve a maximal violation of $S_{\max}^{\mathrm{disp}}\approx3.78$, it is expected that considering only a few measurements settings ($K=4$ as shown in Fig.~\ref{F:displacement-free-ref-frame}) suffices to guarantee a violation for any rotation angle. By contrast, homodyne measurements attain a lower maximum ($S_{\max}^{\mathrm{hom}}\approx3.11$), reflecting on the necessity of a larger number of available settings to ensure violations across all reference-frame orientations. 

\section{Discussion and Outlook}\label{Sec:final-remarks}

In this paper, we studied semi–device–independent randomness certification in prepare‑and‑measure scenarios on continuous‑variable platforms, enforcing the dimension assumption by restricting preparations to the two-dimensional Fock subspace. We introduced a simple discretization of homodyne outcomes and demonstrated that violations of dimension witnesses are achievable using only homodyne measurements. Since homodyne detection can be performed with efficiencies approaching unity, our approach paves the way for robust and experimentally less demanding loophole-free tests. Further, we derived a closed‑form witness‑to‑randomness map for the tilted witness $S_3(w)$, which yields analytic lower bounds on extractable min‑entropy per use. We optimized and compared homodyne‑only, displacement‑only, and hybrid configurations in the $(3,2,2,2)$, $(4,2,2,2)$ and $(3,3,2,2)$ scenarios. Aditionally, for the three‑measurement case $(3,3,2,2)$ we established previously unknown quantum maxima, $S_{33,1}^{\mathrm Q}=3\sqrt{3}$ and $S_{33,2}^{\mathrm Q}=6$. From the observed violations, we certified positive randomness up to $H_\infty\approx0.382$ bits for optimized $S_3(w)$, and $H_\infty=0.342$ and $0.263$ bits at the quantum maxima of $S_{33,1}$ and $S_{33,2}$, respectively. For $S_3(w)$, we also derived an analytical bound on the min entropy, providing a  device-independent and method‑agnostic guarantee. Robustness analyses incorporating amplitude‑damping losses and unknown reference‑frame phases show that violations, and thus certified randomness, persist well below ideal conditions. To our knowledge, we are the first to perform a reference‑frame analysis in the PAM scenario, which had previously been restricted to Bell scenarios~\cite{shadbolt2012guaranteed,brask2013testing}. Notably, four appropriately chosen displacement settings suffice to guarantee a violation for any relative phase, and including at least one homodyne setting further lowers the critical detection-efficiency threshold across witnesses. Together, these results outline a practical and scalable approach to SDI randomness certification in continuous‑variable optics using only standard homodyne and displacement measurements.

Our results open the door to several straightforward directions for future work. First, the loss model can be enriched to include dark counts, mode mismatch, and phase-drift dynamics. Studying these jointly with optimized binning strategies (allowing non-interval sets $A^\pm$) may further lower $\eta_{\mathrm{crt}}$ and increase the certified $H_\infty$. Second, beyond on/off detection and standard homodyne, incorporating photon-number-resolving detectors, heterodyne schemes, or modest non-Gaussian resources (e.g., heralded operations) could boost witness violations within the same two-dimensional subspace, or allow controlled extensions to larger effective dimensions while preserving SDI assumptions. Finally, a more challenging but impactful path is to develop practical protocols: a finite-statistics, composable analysis --e.g., via entropy accumulation-- that translates per-trial $H_\infty$ into secure bit or key rates under realistic data sizes would bridge this framework to deployable QRNGs.

While our analysis restricts the communicated system to the lowest two-level Fock subspace, the same ideas extend naturally to other continuous-variable regimes. A key complementary SDI route replaces this hard dimension assumption with a physically motivated energy constraint~\cite{VanHimbeeck2017,VanHimbeeck2019Energy}, which underlies several practical homodyne/heterodyne QRNG protocols~\cite{Rusca2019SelfTesting,Rusca2020Homodyne,Avesani2021Heterodyne,Wang2023Homodyne}. For instance, one may use a single-mode squeezed state as the carrier and certify randomness via an energy witness~\cite{Roch2025}, exploring whether Gaussian measurements alone can suffice. On the implementation side, a pulsed optical testbed---with weak coherent-state preparations truncated to the $\{\ket{0},\ket{1}\}$ subspace via the quantum scissors protocol~\cite{pegg1998optical,babichev2003quantum} and high-efficiency detection---offers a minimal, practical platform for randomness generation and a direct test of our SDI certification, bridging the qubit-encoded CV model to scalable photonic QRNGs.

Recent work has also emphasized the role of entanglement in PAM scenarios. In dimension-bounded setups, entanglement between sender and receiver can generate correlations inaccessible to unassisted models~\cite{Tavakoli2021,Pauwels2022,RochTavakoli2024,Pawlowski2010,moreno2021semi}. In energy-constrained PAM, entanglement with an adversary may reduce certifiable randomness~\cite{DAvino2025}, although security largely persists in low-energy optical implementations~\cite{RochCarceller2025}. These results clarify when entanglement acts as a resource or a limitation and motivate future work to incorporate such effects into CV SDI analyses.

\begin{acknowledgments} 
This work was supported by the Simons Foundation (Grant No. 1023171, R.C.), the Brazilian National Council for Scientific and Technological Development (CNPq, Grants No. 307295/2020-6, No.403181/2024-0, and 301687/2025-0), the Financiadora de Estudos e Projetos (Grant No. 1699/24 IIF-FINEP), and the Coordenação de Aperfeiçoamento de Pessoal de Nível Superior, Brasil (CAPES) Finance Code 001 and a guest professorship from the Otto M\o nsted Foundation.  Additional funds were provided by the project ``Certificação de Aleatoriedade Quântica'' supported by QuIIN - Quantum Industrial Innovation, EMBRAPII CIMATEC Competence Center in Quantum Technologies, with financial resources from the PPI IoT/Manufatura 4.0 of the MCTI grant number 053/2023, signed with EMBRAPII. AOJ acknowledges financial support from the  EU Horizon Europe (QSNP, grant no. 101114043) \& the Danish National Research Foundation grant bigQ (DNRF 142).  
\end{acknowledgments}

\section*{Data Availability}
The data that support the findings of this article are openly
available~\cite{sena2025sdi}.

\bibliographystyle{apsrev4-2}
\bibliography{biblio}

\clearpage 

\appendix
\section{Optimal strategies for optimal witnesses values}\label{Sec:app-optimal-strategies}
\label{app: opt_params}
In this appendix, we list the parameters yielding the optimal violations of the dimension witnesses studied in this work. We consider states of the form in Eq.~\eqref{eq: rho_bloch_parameterized} and homodyne or displacement POVMs as in Eqs.~\eqref{eq: homodyne_projector} and \eqref{eq: displacement_projector}. Following the parameterization of Sec.~\ref{sec: resultsA}, homodyne settings use a binning \(A^+ = [a,b]\) (with \(A^- = \mathbb{R}\setminus A^+\)), with \(a,b \in [-5,5]\) and quadrature angle \(0 \le \theta \le \pi\). Displacement measurements use \(\alpha = r e^{i\phi}\) with \(0 \le r \le 1\) and \(0 \le \phi < 2\pi\). Optimal strategies for each scenario are listed in Tables~\ref{tab:params_3222}, \ref{tab:params_4222}, and \ref{tab:params3322}. Note that these optimal strategies are not unique: different choices of state and measurement parameters may lead to the same value of the witness, for instance, due to symmetries in the Bloch parametrization or equivalent relabelings of outcomes and inputs.

% Editar tabelas ainda
\begin{table}[h]
\centering
\caption{Optimal parameters for the $(3,2,2,2)$ scenario for $S_3$: state angles $\alpha_x$, phases $\eta_x$, and measurement parameters $(a,b,\theta)_y$ for homodyne or $(r,\phi)_y$ for displacement.}
\label{tab:params_3222}
\small
\setlength{\tabcolsep}{6pt}
\renewcommand{\arraystretch}{1.25}
\begin{tabular}{c|l}
\hline\hline
Setup & Parameters \\ 
\hline

HH &
$\alpha_{1\text{--}3} = [1.4681,\ 1.4681,\ 1.7188]$ \\
& $\eta_{1\text{--}3} = [2.3339,\ 3.9492,\ 6.2832]$ \\
& $(a,b,\theta)_1 = [-4.6243,\ 0.2109,\ 0.0000]$\\
& $(a,b,\theta)_2 = [-0.0000,\ 3.7764,\ 1.5707]$\\

\hline
DH &
$\alpha_{1\text{--}3} = [0.6734,\ 0.6734,\ 3.1416]$ \\
& $\eta_{1\text{--}3} = [5.5820,\ 2.4404,\ 0.5144]$ \\
& $(r,\phi)_1 = [0.0000,\ 2.5857]$ \\
& $(a,b,\theta)_2 = [-4.9984, 0.0000, 2.4404]$, \\

\hline
HD &
$\alpha_{1\text{--}3} = [0.4936,\ 2.3271,\ 1.7320]$ \\
& $\eta_{1\text{--}3} = [1.5682,\ 1.5404,\ 4.6931]$ \\
& $(a,b,\theta)_1 = [-0.2299,\ 4.1153,\ 1.5515]$ \\
& $(r,\phi)_2 = [0.0800,\ 1.6542]$ \\

\hline
DD &
$\alpha_{1\text{--}3} = [0.0926,\ 1.4679,\ 2.5297]$ \\
& $\eta_{1\text{--}3} = [0.0000,\ 3.1415,\ 6.2832]$ \\
& $(r,\phi)_1 = [0.3159,\ 3.1415]$ \\
& $(r,\phi)_2 = [0.4263,\ 6.2832]$ \\
\hline\hline
\end{tabular}
\end{table}

\begin{table}[tbp]
\centering
\caption{Optimal parameters for the $(4,2,2,2)$ scenario for $S_4$: state angles $\alpha_x$, phases $\eta_x$, and measurement parameters $(a,b,\theta)_y$ for homodyne or $(r,\phi)_y$ for displacement.}
\label{tab:params_4222}
\small
\setlength{\tabcolsep}{6pt}
\renewcommand{\arraystretch}{1.25}
\begin{tabular}{c|l}
\hline\hline
Setup & Parameters \\ 
\hline

HH &
$\alpha_{1\text{--}4} = [1.5706,\ 1.5709,\ 1.5707,\ 1.5706]$ \\[2pt]
& $\eta_{1\text{--}4} = [0.6903,\ 2.2612,\ 5.4028,\ 3.8320]$ \\[2pt]
& $(a,b,\theta)_1 = [0.0000,\ 3.9969,\ 1.4758]$ \\[2pt]
& $(a,b,\theta)_2 = [-4.7507,\ 0.0000,\ 3.0464]$ \\[4pt]

\hline
HD &
$\alpha_{1\text{--}4} = [0.6727,\ 2.4674,\ 0.6737,\ 2.4688]$ \\[2pt]
& $\eta_{1\text{--}4} = [2.6846,\ 2.6932,\ 5.8344,\ 5.8265]$ \\[2pt]
& $(a,b,\theta)_1 = [-0.0002,\ 4.0793,\ 2.6888]$ \\[2pt]
& $(r,\phi)_2 = [0.0017,\ 5.4447]$ \\[4pt]

\hline
DD &
$\alpha_{1\text{--}4} = [0.0000,\ 1.5705,\ 1.5708,\ 3.1414]$ \\[2pt]
& $\eta_{1\text{--}4} = [4.2027,\ 2.5052,\ 5.6471,\ 5.2347]$ \\[2pt]
& $(r,\phi)_1 = [0.3820,\ 3.7780]$ \\[2pt]
& $(r,\phi)_2 = [0.3820,\ 0.6361]$ \\[4pt]

\hline\hline
\end{tabular}
\end{table}

\begin{table}[tbp]
\centering
\caption{Optimal parameters for the $(3,3,2,2)$ scenario for $S_{33,1}$ and $S_{33,2}$: state angles $\alpha_x$, phases $\eta_x$, and measurement parameters $(a,b,\theta)_y$ for homodyne or $(r,\phi)_y$ for displacement.}
\label{tab:params3322}
\small
\setlength{\tabcolsep}{6pt}
\renewcommand{\arraystretch}{1.25}
\begin{tabular}{c|c|l}
\hline\hline
Witness & Setup & Parameters \\ 
\hline

\multirow{2}{*}{$S_{33,1}$} 
& HHH &
$\alpha_{1\text{--}3} = [1.5708,\ 1.5708,\ 1.5708]$ \\[2pt]
& & $\eta_{1\text{--}3} = [0.8635,\ 2.9580,\ 5.0523]$ \\[2pt]
& & $(a,b,\theta)_1 = [0.0000,\ 4.5036,\ 1.3870]$ \\[2pt]
& & $(a,b,\theta)_2 = [0.0000,\ 4.5840,\ 0.3399]$ \\[2pt]
& & $(a,b,\theta)_3 = [0.0000,\ 3.8455,\ 2.4344]$ \\[4pt]

& DDD &
$\alpha_{1\text{--}3} = [0.4426,\ 1.5707,\ 2.6992]$ \\[2pt]
& & $\eta_{1\text{--}3} = [1.6888,\ 4.8305,\ 1.6891]$ \\[2pt]
& & $(r,\phi)_1 = [0.0000,\ 1.8640]$ \\[2pt]
& & $(r,\phi)_2 = [0.4807,\ 4.5944]$ \\[2pt]
& & $(r,\phi)_3 = [0.4806,\ 1.4526]$ \\[4pt]

\hline

\multirow{2}{*}{$S_{33,2}$} 
& HHH &
$\alpha_{1\text{--}3} = [1.4145,\ 1.5754,\ 1.5752]$ \\[2pt]
& & $\eta_{1\text{--}3} = [2.5416,\ 0.4284,\ 4.6551]$ \\[2pt]
& & $(a,b,\theta)_1 = [-0.2213,\ 3.7027,\ 1.4940]$ \\[2pt]
& & $(a,b,\theta)_2 = [-3.2934,\ 0.2213,\ 0.4477]$ \\[2pt]
& & $(a,b,\theta)_3 = [0.0000,\ 2.2903,\ 2.5418]$ \\[4pt]

& DDD &
$\alpha_{1\text{--}3} = [0.0000,\ 2.1577,\ 2.1577]$ \\[2pt]
& & $\eta_{1\text{--}3} = [2.6398,\ 3.1417,\ 0.0000]$ \\[2pt]
& & $(r,\phi)_1 = [0.4680,\ 3.1415]$ \\[2pt]
& & $(r,\phi)_2 = [0.4680,\ 0.0000]$ \\[2pt]
& & $(r,\phi)_3 = [0.0000,\ 3.7669]$ \\[4pt]

\hline\hline
\end{tabular}
\end{table}

\section{Derivation of the analytical upper bound on the guessing probability}\label{Sec:app-analytical-upper-bound}
\label{app: analytical_upper_bound}
Consider the probabilities of the form of Eq.~\eqref{eq: uniform_avg_pguess}, such that
\begin{align}
\overline{p}_{\mathrm{guess}}^{(u) \lambda} &= \frac{1}{n_xn_y}\sum_{x,y} \max_{b} \, p(b|x,y,\lambda)
\end{align}
with $n_y =2$ and arbitrary $n_x$. The shared randomness $\lambda$  determines the behavior of the measurement and preparation devices in each round of the experiment. We assume it is distributed by $q(\lambda)$.
Consider the bloch decomposition of preparation and measurements:
\begin{equation}
    \rho^\lambda_x =\frac{1}{2}(\iden_2 +  \vec{r}^\lambda_x\cdot\vec{\sigma}) \quad\quad M^\lambda_{b|y} =\frac{1}{2}(\iden_2 + (-1)^b \vec{n}^\lambda_y\cdot\vec{\sigma}).\label{eq: states_meas_bv}
\end{equation}
Then the probability reads
\begin{equation}
    p(b|x,y,\lambda) = \frac{1}{2}(1+(-1)^b \vec{r}^\lambda_x\cdot\vec{n}^\lambda_y),
\end{equation}
Maximizing over $b$ and summing over the two settings we get
\begin{equation}
    \sum_{y=0}^1\max_bp(b|x,y,\lambda)= 1 + \frac{1}{2}(|\vec{r}^\lambda_x\cdot\vec{n}^\lambda_1| + |\vec{r}^\lambda_x\cdot\vec{n}^\lambda_2|).
\end{equation}
Let $\vec{n}_1^\lambda$ and $\vec{n}_2^\lambda$ be vectors in a plane separated by an angle $\theta^\lambda$, and let $\phi_x$ be the angle of $\vec{r}_x^\lambda$ with respect to the bisector of $\vec{n}_1^\lambda$ and $\vec{n}_2^\lambda$. Without loss of generality, we place $\vec{n}_1^\lambda$ and $\vec{n}_2^\lambda$ at angles $\pm \theta^\lambda/2$ with respect to the bisector, so that
\begin{align}
    \vec{r}^\lambda_x\cdot\vec{n}^\lambda_1 &= \cos(\phi_x-\theta^\lambda/2) \| \vec{r}^\lambda_x\|_2\|\vec{n}^\lambda_1\|_2, \nonumber\\
    \vec{r}^\lambda_x\cdot\vec{n}^\lambda_2 &= \cos(\phi_x+\theta^\lambda/2) \| \vec{r}^\lambda_x\|_2\|\vec{n}^\lambda_2\|_2.
\end{align}
Then, 
\begin{align}
    \sum_{y=0}^1\max_bp(b|x,y,\lambda)= 1 &+  \frac{1}{2}\bigg(|\cos(\phi_x-\theta^\lambda/2)|\, \| \vec{r}^\lambda_x\|_2\|\vec{n}^\lambda_1\|_2 \nonumber\\&+ |\cos(\phi_x+\theta^\lambda/2)|\, \| \vec{r}^\lambda_x\|_2\|\vec{n}^\lambda_2\|_2\bigg).
\end{align}
The maximum of the above quantity is attained for pure preparations and projective measurements: $| \vec{r}^\lambda_x\|_2=\|\vec{n}^\lambda_y\|_2=1$ for all $x,y$. Also the optimal strategy is to take $\vec{r}^\lambda_x$ with positive projection onto $\vec{n}^\lambda_1$ and $\vec{n}^\lambda_2$, \emph{i.e.} $\cos(\phi_x-\theta/2)>0$ and $\cos(\phi_x+\theta/2)>0$.  Then, maximizing over all $x$, we have 
\begin{align}
     &\max_x\sum_{y=0}^1\max_bp(b|x,y,\lambda) \leq\nonumber\\
     &\leq1+\frac{1}{2}\max_x\{\cos(\phi_x-\theta^\lambda/2) + \cos(\phi_x+\theta^\lambda/2)\}\nonumber\\
     &=1+\frac{1}{2}\max_x\{2\cos(\phi_x)\cos(\theta^\lambda/2)\}\nonumber\\
     &\leq1+\cos(\theta^\lambda/2).
\end{align}
Now, observe that 
\begin{align}
\sum_{x,y} \max_{b} \, p(b|x,y,\lambda) &\leq n_x \sum_{y} \max_{b,x}\, p(b|x,y,\lambda),
\end{align}
so putting all together we have
\begin{equation}
   \overline{p}_{\mathrm{guess}}^{(u) \lambda}  \leq  \frac{1}{n_y} \bigg(1+\cos(\theta^\lambda/2)\bigg).\label{eq: A_bound_pguess}
\end{equation}

Consider now the value of the tilted $S_3$ witness defined in Eq.~\eqref{eq: s3_tilted} as
\begin{equation}S^\lambda_3(w) = w(E^\lambda_{11}+E^\lambda_{21}-E^\lambda_{31}) + (1-w)(E^\lambda_{12}-E^\lambda_{22}).\end{equation}
Note that by expressing the states and measurements as in Eq.~(\ref{eq: states_meas_bv}) the correlator reduces to $E_{xy} = p(0|xy)-p(1|xy)= \vec{r}_x\cdot \vec{n}_y$, so we can rewrite
\begin{align}
    S_3(w) &= \vec{r}_1\cdot(w(\vec{n}^\lambda_1-\vec{n}^\lambda_2)+\vec{n}^\lambda_2) \nonumber\\&+ \vec{r}_2\cdot(w(\vec{n}^\lambda_1+\vec{n}^\lambda_2)-\vec{n}^\lambda_2) -w\vec{r}_3\cdot\vec{n}^\lambda_1, 
\end{align}
where the value of $S_3(w)$ is maximized if $\vec{r_1} = (w(\vec{n}^\lambda_1-\vec{n}^\lambda_2)+\vec{n}^\lambda_2)/\|w(\vec{n}^\lambda_1-\vec{n}^\lambda_2) + \vec{n}^\lambda_2\|_2$, $\vec{r_2} = (w(\vec{n}^\lambda_1+\vec{n}^\lambda_2)-\vec{n}_2)/\|w(\vec{n}^\lambda_1+\vec{n}^\lambda_2) - \vec{n}^\lambda_2\|_2$, and $\vec{r}_3 = - \vec{n}^\lambda_1/ \|\vec{n}^\lambda_1\|_2$. This implies that 
\begin{equation}
    S_3(w)\leq\|\vec{u}_{+}\|_2 +\|\vec{u}_{-}\|_2 + w,
\end{equation}
where  $\vec{u}_{\pm} = w\vec{n}^\lambda_1 \pm (1-w)\vec{n}^\lambda_2$.
Now see that 
\begin{equation}
    \|\vec{u}_{\pm}\|_2^2 = w^2\pm2w(1-w)\vec{n}^\lambda_1\cdot\vec{n}^\lambda_2 + (1-w)^2,
\end{equation}
recalling that the angle between the two measurement vectors is $\theta^\lambda$ and that $S_3(w)$ is maximized when $\|\vec{n}^\lambda_1\|_2 = \|\vec{n}^\lambda_2\|_2 = 1$, then 
\begin{equation}
    \|\vec{u}_{\pm}\|_2^2 = w^2\pm2w(1-w)\cos(\theta^\lambda) + (1-w)^2.
\end{equation}
By defining $A= A(w) =  w^2 + (1-w)^2$ and $B = B(w) = 2w(1-w)$, then one gets that 
$S_3(w) = w + \sqrt{A + B\cos(\theta^\lambda)}  + \sqrt{A - B\cos(\theta^\lambda)}$.
From the above equation, one can isolate $\cos(\theta^\lambda)$, obtaining
\begin{equation}
    |\cos(\theta^\lambda)| = \sqrt{\frac{A^2-[(S^\lambda_3(w)-w)^2/2-A]^2}{B^2}},
\end{equation}
for $w\neq0,1$.
From this equation and using the result~(\ref{eq: A_bound_pguess}), we get that 
\begin{align}
     \overline{p}_{\mathrm{guess}}^{(u) \lambda}  &\leq  \frac{1}{2} (1+\cos(\theta^\lambda/2))\nonumber\\
     &\leq\frac{1}{2} (1+|\cos(\theta^\lambda/2)|) \nonumber\\
     &\leq\frac{1}{2} \big(1+\sqrt{\frac{1+|\cos(\theta^\lambda)|}{2}}\big)\nonumber\\
     &=\frac{1}{2}+\frac{1}{2}\sqrt{\frac{1}2 + \frac{A^2-[(S^\lambda_3(w)-w)^2/2-A]^2}{2B^2}}.
\end{align}
Finally by noting that  $S_3(w) = \int S_3^\lambda(w)q(\lambda)d\lambda$, the guess probability is 
\begin{equation}
   \overline{p}_{\mathrm{guess}}^{(u)}  = \int \overline{p}_{\mathrm{guess}}^{(u) \lambda} q(\lambda)d\lambda \leq \int F(S_3^\lambda)q(\lambda)d\lambda,
\end{equation}
where 
\begin{align}
    F[S_3^\lambda(w)] = \frac{1}{2} +\frac{1}{2}\sqrt{\frac{1}2 + \frac{A^2-[(S^\lambda_3(w)-w)^2/2-A]^2}{2B^2}}.
\end{align}
Jensen's inequality says that if $X$ is a random variable and $\varphi$ is a concave function then $\varphi(E(X))\geq E(\varphi(X))$. By setting $\varphi = F$, the concavity of the function $F$ implies that
\begin{equation}
    \overline{p}_{\mathrm{guess}}^{(u)} \leq F[S_3(w)].
\end{equation}

\section{Quantum bounds for the witnesses of the (3,3,2,2) scenario}
\label{app: analytical_Q_bounds}
In this appendix we derive the maximal quantum values attainable by qubit systems for the witnesses $S_{33,1}$ and $S_{33,2}$ in the $(3,3,2,2)$ scenario.

\subsection{Quantum bound of $S_{33,1}$}
The witness
\begin{align}
S_{33,1} = E_{11} + E_{12} - E_{22} + E_{23} - E_{31} - E_{33}
\end{align}
can be written as
\begin{align}
S_{33,1}
= \vec r_1 \cdot (\vec m_1+\vec m_2)
+ \vec r_2 \cdot (\vec m_3-\vec m_2)
- \vec r_3 \cdot (\vec m_1+\vec m_3),
\end{align}
where we parametrized the correlations as \(E_{xy}=\vec r_x\cdot\vec m_y\), with
\(\vec r_x\) and \(\vec m_y\) unit Bloch vectors describing the prepared states
and the measurement directions, respectively. For fixed measurements, the witness is maximized by aligning each preparation Bloch vector with the corresponding vector appearing in the scalar product, yielding
\begin{align}
S_{33,1}
\le |\vec m_1+\vec m_2|
+|\vec m_3-\vec m_2|
+|\vec m_1+\vec m_3|.
\end{align}
Let $\theta_{ij}$ denote the angle between the unit Bloch vectors $\vec m_i$ and $\vec m_j$. Using
\begin{align}
|\vec m_i \pm \vec m_j|
= \sqrt{(\vec m_i \pm \vec m_j)\cdot(\vec m_i \pm \vec m_j)}
= \sqrt{2 \pm 2\cos\theta_{ij}},
\end{align}
the expression to be maximized becomes
\begin{align}
\sqrt{2+2\cos\theta_{12}}
+\sqrt{2-2\cos\theta_{23}}
+\sqrt{2+2\cos\theta_{13}} .
\end{align}
Without loss of generality the optimal measurement Bloch vectors can
be taken to lie in a common plane. Let $\alpha_i$ denote the planar
angle of $\vec m_i$, and fix $\alpha_1=0$ (a global rotation does not
change the value of the witness). Defining \( \theta_{12} = \alpha_2 \), \( \theta_{13} = \alpha_3 \), and \( \theta_{23} = \alpha_2 - \alpha_3 \), the problem reduces to maximizing
\begin{align}
f(\alpha_2,\alpha_3)
=
\sqrt{2+2\cos\alpha_2}\notag
&+
\sqrt{2-2\cos(\alpha_2-\alpha_3)}
\\&+
\sqrt{2+2\cos\alpha_3}.
\end{align}
The maximum is obtained for a configuration symmetric with respect to
$\vec m_1$, namely \(\alpha_2 = \frac{\pi}{3}\) and \(\alpha_3 = -\frac{\pi}{3}\), which gives $\cos\theta_{12}=\cos\theta_{13}=\frac{1}{2}$ and $\cos\theta_{23}=-\frac{1}{2}$. Substituting into the norms yields
\begin{align}
|\vec m_1+\vec m_2|
=
|\vec m_3-\vec m_2|
=
|\vec m_1+\vec m_3|
=
\sqrt{3}.
\end{align}
Therefore
\begin{align}
S_{33,1} \le 3\sqrt{3},
\end{align}
and hence \(S_{33,1}^{Q}=3\sqrt{3}.\)

\subsection{Quantum bound of $S_{33,2}$}
The witness reads
\begin{align}
S_{33,2}
&=
E_{11}+E_{12}+E_{13} +E_{21}\notag\\
&\quad -E_{22}-E_{23}-E_{31}+E_{32}-E_{33}.
\end{align}
Using the Bloch representation one can write
\begin{align}
S_{33,2}
&=
\vec r_1\!\cdot\!(\vec m_1+\vec m_2+\vec m_3) \notag +\vec r_2\!\cdot\!(\vec m_1-\vec m_2-\vec m_3) \\
&\quad +\vec r_3\!\cdot\!(-\vec m_1+\vec m_2-\vec m_3).
\end{align}
For fixed measurement directions the expression is maximized by
aligning each preparation vector with the vector multiplying it,
which yields
\begin{align}
S_{33,2}
&\le
|\vec m_1+\vec m_2+\vec m_3| + \notag
|\vec m_1-\vec m_2-\vec m_3| \\
&\quad +
|-\vec m_1+\vec m_2-\vec m_3|.
\end{align}
Let $s_{ij}=\vec m_i\cdot\vec m_j$. Expanding the squared norms gives
\begin{align}
|\vec m_1+\vec m_2+\vec m_3|^2
&=3+2(s_{12}+s_{13}+s_{23}),
\\
|\vec m_1-\vec m_2-\vec m_3|^2
&=3-2(s_{12}+s_{13}-s_{23}),
\\
|-\vec m_1+\vec m_2-\vec m_3|^2
&=3-2(s_{12}-s_{13}+s_{23}).
\end{align}
The expression to be maximized depends on $s_{13}$ and $s_{23}$
only through symmetric combinations of these variables.
Therefore if a configuration $(s_{12},s_{13},s_{23})$ is optimal,
the configuration obtained by exchanging $s_{13}$ and $s_{23}$
achieves the same value. Consequently one may choose an optimal configuration on the
symmetric line $s_{13}=s_{23}$, which we parameterize as
\begin{align}
s_{13}=s_{23}=t,
\qquad
s_{12}=u .
\end{align}
Substituting these relations gives
\begin{align}
|\vec m_1+\vec m_2+\vec m_3|^2 &= 3+2(u+2t),\\
|\vec m_1-\vec m_2-\vec m_3|^2 &= 3-2u,\\
|-\vec m_1+\vec m_2-\vec m_3|^2 &= 3-2u.
\end{align}
The parameters $t$ and $u$ cannot be chosen arbitrarily, since the
scalar products must correspond to a valid set of unit vectors
$\vec m_1,\vec m_2,\vec m_3$. Equivalently, their Gram matrix
\begin{align}
G=
\begin{pmatrix}
1 & u & t\\
u & 1 & t\\
t & t & 1
\end{pmatrix}
\end{align}
must be positive semidefinite, i.e., $G\succeq 0$. The determinant condition $\det G\ge0$
then implies
\begin{align}
2t^2-1 \le u \le 1 ,
\end{align}
where the upper bound follows from the inner product of two unit vectors. Maximizing the sum of norms under this constraint yields the optimal
values $t=\frac12$ and $u=-\frac12$, for which
\begin{align}
|\vec m_1+\vec m_2+\vec m_3|
=
|\vec m_1-\vec m_2-\vec m_3|
=
|-\vec m_1+\vec m_2-\vec m_3|
=
2 .
\end{align}
Hence
\begin{align}
S_{33,2}\le 6 .
\end{align}
This value is attained by coplanar measurement directions separated
by angles $0$, $2\pi/3$ and $\pi/3$, for instance
$\vec m_1=(1,0,0)$,
$\vec m_2=\left(-\tfrac12,\tfrac{\sqrt3}{2},0\right)$,
$\vec m_3=\left(\tfrac12,\tfrac{\sqrt3}{2},0\right)$,
with preparations aligned with
$\vec m_1+\vec m_2+\vec m_3$,
$\vec m_1-\vec m_2-\vec m_3$,
and $-\vec m_1+\vec m_2-\vec m_3$,
which saturates the bound $S_{33,2}^Q = 6$.
\end{document}